\documentclass[prd,aps,preprint,superscriptaddress]{revtex4-1} 

\usepackage[utf8]{inputenc}
\usepackage[english]{babel}
\usepackage{amsmath}
\usepackage{amssymb}
\usepackage{amsfonts}
\usepackage{mathrsfs}
\usepackage{graphicx}
\usepackage{xcolor}
\usepackage{bm}
\usepackage{subcaption}
\usepackage[abs]{overpic}
\usepackage{pifont}

\usepackage{changepage}
\usepackage{mathtools}

\newcommand{\new}[1]{#1}

\newcommand{\tunedvs}{3pt}

\newcommand{\lp}{{\overline{\varphi}}{}}
\renewcommand{\ll}{\overline{\overline{\varphi}}{}}
\renewcommand{\lll}{\overline{\overline{\overline{\varphi}}}{}}
\newcommand{\llll}{\overline{\overline{\overline{\overline{\varphi}}}}{}}

\newcommand{\G}{\overline{g}{}}
\newcommand{\GG}{\overline{\overline{g}}{}}
\newcommand{\GGG}{\overline{\overline{\overline{g}}}{}}
\newcommand{\g}{\widehat{\overline{g}}{}}
\renewcommand{\gg}{\widehat{\overline{\overline{g}}}{}}
\renewcommand{\ggg}{\widehat{\overline{\overline{\overline{g}}}}{}}

\renewcommand{\deg}{\operatorname{deg}}

\newcommand{\col}[1]{\textbf{\textcolor{violet}{(C#1)}}}

\newcommand{\I}{\mathcal{I}}

\newcommand{\coeffs}[1]{{\textcolor{blue}{#1}}} 
\newcommand{\straight}[1]{{\textcolor{red}{#1}}}
\newcommand{\tilted}[1]{{\textcolor{red}{#1}}} 
\newcommand{\init}[1]{{\textcolor{red}{#1}}} 

\newcommand{\CC}[3]{\coeffs{C_{#1 : #2}{}^{#3}}}
\newcommand{\Coeff}[1]{\coeffs{C_{#1}}}

\begin{document}
\title{Gravitational closure of matter field equations}
\author{Maximilian D\"ull}
\affiliation{Universit\"at Heidelberg, Zentrum f\"ur Astronomie, Philosophenweg 12, 69120 Heidelberg}
\author{Frederic P. Schuller\footnote{Corresponding author. Electronic address fps@aei.mpg.de}} 
\affiliation{Friedrich-Alexander Universit\"at Erlangen-N\"urnberg, Department Physik, Staudtstr. 7, 91058 Erlangen, Germany}
\author{Nadine Stritzelberger}
\affiliation{University of Cambridge, Newnham College, Sidgwick Avenue, Cambridge CB3 9DF, United Kingdom}
\affiliation{University of Waterloo, Department of Applied Mathematics, Waterloo N2L 3G1, Ontario, Canada}
\author{Florian Wolz${}^2$}

\begin{abstract}
\noindent 
\new{The requirement that both the matter and the geometry of a spacetime canonically evolve together, starting and ending on shared Cauchy surfaces and independently of the intermediate foliation, leaves one with little choice for diffeomorphism-invariant gravitational dynamics that can 
provide consistent evolution equations to the coefficients of 
a given system of matter field equations. Concretely, we show how starting from any linear local matter field equations whose principal polynomial satisfies three physicality conditions, one may calculate coefficient functions which then enter an otherwise immutable set of countably many linear homogeneous partial differential equations. Any solution of these so-called gravitational closure equations then provides a Lagrangian density for any type of tensorial geometry that features ultralocally in the initially specified matter Lagrangian density. Thus the given system of matter field equations is indeed closed by the so obtained gravitational equations. In contrast to previous work, we build the theory on a suitable associated bundle encoding the canonical configuration degrees of freedom, which allows to include necessary constraints on the geometry in practically tractable fashion. By virtue of the presented mechanism, one thus can practically calculate, rather than having to postulate, the gravitational theory that is required by specific matter field dynamics. For the special case of standard model matter one obtains general relativity.}

\end{abstract}

\newpage
\maketitle
\tableofcontents
\newpage
\section{Introduction}
There remains an uncomfortable arbitrariness in the construction of modified gravity models, which even plagues the proposals that heed the currently known theoretical and observational constraints \cite{Clifton,Joyce,Koyama}. Since a finite number of experiments will not be able to discriminate against an infinity of models, bona fide physical input must likely be injected into the construction beforehand, instead of being left to discriminate against theories only afterwards. 

In this article, we argue that such genuine physical input, which promises to effect a reduction of the current infinite ambiguity toward a finite one, is provided by first prescribing the matter dynamics on a spacetime. \new{The dynamics of the underpinning spacetime geometry are then shown to follow from the matter dynamics, essentially by a sufficiently precise requirement of common canonical evolution. Note that the thus revealed dependence of the resulting gravitational dynamics on previously specified matter dynamics implies that there is no one-size-fits-all gravity theory for a given geometry, which would apply independently of what we know or discover about matter. This is because the matter dynamics crucially determine the kinematical meaning of their geometric background, and it is precisely this information that directly funnels into the structure of the gravitational dynamics. Through this mechanism, any new insight into the nature of matter may yield new information about gravity. 
Not too bad a perspective in the first place, in face of having detailed knowledge of the fundamental dynamics for only 4.6\% of the matter and energy in the universe.} And not a new perspective either, considering that it was the dynamics of matter, namely the classical electromagnetic field, that led Einstein to the identification and kinematical interpretation of Lorentzian geometries and finally the field equations for their dynamics. 

\new{Precisely, we show the following. For any diffeomorphism invariant matter action whose integrand depends locally on some tensorial matter field $A$ and ultralocally on a geometric background described by some tensor field $G$ of arbitrary valence,  
\begin{equation}\label{intromatter}
S_\text{\tiny matter}[A;G) 
= \int \mathrm d^4 x \,\mathscr{L}_\textrm{\tiny matter}(A(x),\partial A(x), \dots, \partial^\textrm{\tiny finite}\!A(x); G(x))\,,
\end{equation}
and which satisfies the three {\it matter conditions} detailed in section \ref{subsec_pdr}, we show how to calculate four geometry-dependent coefficients $E^A{}_\mu$, $F^A{}_\mu{}^{\nu}$, $M^{B\mu}{}$ and $p^{\alpha\beta}$ that enter the {\it gravitational closure equations}, which is the countable set of linear homogeneous partial differential equations displayed on the final two landscape pages of this paper. Their solution then, in turn, provides the closure 
\begin{equation}\label{Sclosed}
S_\text{\tiny closed}[A;G] 
= S_\text{\tiny matter}[A;G) + \int \mathrm d^4 x\, \mathscr{L}_\textrm{\tiny geometry}(G(x),\partial G(x), \dots,\partial^\textrm{\tiny finite}G(x))\,
\end{equation}
of the given matter field dynamics by inclusion of gravitational dynamics whose canonical version satisfies the two {\it embedding properties} laid out in section \ref{subsec_HKT}. } 

\new{Indeed, only after addition of the thus calculated, rather than stipulated gravitational Lagrangian density $\mathscr{L}_\textrm{\tiny geometry}$ is there a dynamically closed theory. For now varying the complete action (\ref{Sclosed}) with respect to the matter field $A$ still yields the matter field equations, while variation with respect to on the previously undetermined background $G$ provides additional equations of motion for the previously unspecified geometry that are sourced by the given matter. Thus a closed system of equations that determines all unknowns, up to only gauge ambiguities, is obtained.

While a detailed discussion of the above-mentioned matter conditions and embedding properties needs to be deferred to the said sections, it is probably worth to briefly hint at their contents. 
First, all three matter conditions are actually conditions on the so-called principal polynomial of the corresponding field equations. Classically they correspond, in turn, to: the existence of an initial value formulation for the matter field equations; a one-to-one relation between momenta and velocities of massless particles; the requirement of an observer-independent definition of positive particle energy. It is  interesting to note that if one insists on the matter field equations being canonically quantizable, these three properties are directly implied, see \cite{RS} for a concrete demonstration. Second, the two  embedding properties restrict the desired gravitational dynamics such that: geometric data are evolved between any two non-intersecting initial data surfaces in a way that does not depend on the choice of intermediate leaves; the thus generated canonical data are embedded into the spacetime in a consistent way; the resulting theory is invariant under spacetime diffeomorphisms.}

The conceptual and technical developments presented in this article significantly extend and improve the results obtained in \cite{GSWW} in several ways, and spread over the four technical sections of this paper.

\new{First, in section \ref{sec_kinematical}, we show how to derive the principal polynomial of matter field equations even in the presence of gauge symmetries, as is often required in physics. We then list the three matter conditions imposed on the principal polynomial, which need to hold in order for the matter dynamics to effect a complete kinematical interpretation of the geometry that underlies it. Finally, we show in that section how a choice of de-densitization of a primarily obtained principal polynomial density gives rise to a notion of point mass and observer frames. The crucial relevance of this first batch of results lies in the fact that all relevant information in the matter action, as far as the construction of gravitational dynamics for the underlying geometry is concerned, trickles down to the next sections exclusively in form of the triple $(M,G,P)$ consisting of the spacetime manifold $M$, the tensorial geometry $G$ employed in the matter action and the principal polynomial $P$ of the matter field equations that satisfies the three matter conditions.}

Secondly, in section \ref{sec_canonicalkinematics}, we remove a theoretically inexistent, but practically almost prohibitive problem with the application of the results of \cite{GSWW} to kinematical spacetime geometries for which the separation of lapse and shift from true dynamical degrees of freedom imposes non-linear algebraic conditions on the initial data surface geometry. For exactly as in classical mechanics, where the condition that a particle move on a non-linear submanifold of Euclidean space is most effectively dealt with by introduction of generalized coordinates, we also employ generalized tensor field components (corresponding to points in a suitable associated bundle over the spacetime frame bundle), in order to directly deal only with the true degrees of freedom of the theory. The relevant technology, once set up, makes things quite simple.  

Thirdly,  in section \ref{sec_canonicaldynamics}, we now convert the entire constraint algebra for the  
gravitational dynamics into a countable set of linear homogeneous partial differential equations, for whose solution powerful methods are available \cite{SeilerBook}.  Unlike the construction in \cite{GSWW}, this reveals one single and immutable set of equations for the gravitational Lagrangian.  Only the coefficient functions appearing in these partial differential equations vary with the choice of matter dynamics and can now be constructed swiftly according to simple rules. Finally, we show in that section how to completely bypass the Hamiltonian formalism employed in the previous two sections in favor of a Lagrangian spacetime formulation. In particular, we provide a gravitational action functional that depends on the spacetime geometry only, rather than geometric phase space variables. Addition of this spacetime action to the initially provided matter action and subsequent variation with respect to the tensor field $G$ then yields the complete gravitational field equations coupled to matter. 

How truly simple it is to set up the gravitational closure equations for a variety of matter models on different tensorial geometries, is then illustrated by three prototypical examples in section \ref{sec_examples}. In particular, we set up the gravitational closure equations for an instance of standard model matter on a metric manifold, for two scalar fields on a bimetric background and for a refinement of Maxwell theory on a background that does not exclude birefringence a priori. We will, however, not solve the equations for any of these examples here.  

We conclude, in section \ref{sec_conclusions}, by spelling out the impact of our results for both fundamental and phenomenological questions and by pointing out several results we were able to obtain by building on the present article, including explicit perturbative and symmetry-reduced solutions of the gravitational closure equations for phenomenologically interesting or theoretically instructive matter models.

\section{\new{Spacetime kinematics}}\label{sec_kinematical}
This section concisely reviews the constructive steps that need to be performed in order to determine the kinematical interpretation of a tensorial spacetime structure, as it is imprinted by given matter field dynamics on it. The key step in order to extract the kinematical interpretation of a geometry $G$ from specified matter dynamics on it is the calculation of the principal polynomial density $\widetilde P$ of all matter field equations, and we present an explicit method that works also if there is a gauge symmetry. The subsequent imposition of three classical physicality conditions, which however can also be understood as necessary conditions for a canonical quantization of the matter field theory, then restricts the geometry sufficiently to identify massless momenta, observer worldlines, and an observer-independent classification of momenta into such of positive and negative energy. The kinematical structure is then completed by a choice of de-densitization of the principal polynomial density, which allows for a definition of point particle mass and finally of observer frames. The kinematical interpretation of the tensorial spacetime geometry, obtained straight from the stipulated matter dynamics, will flow directly into the gravitational closure equations derived in sections \ref{sec_canonicalkinematics} and \ref{sec_canonicaldynamics}.

\subsection{\new{Several fields for matter and geometry}} \label{subsec_manyfields}
In the interest of avoiding inessential notational clutter, we will present all results of sections \ref{sec_kinematical}, \ref{sec_canonicalkinematics} and \ref{sec_canonicaldynamics} assuming that there is only one single tensor field $A$ describing the matter and only one single tensor field $G$ encoding the underlying geometry, as in the introduction. But all results of these sections straightforwardly generalize to the practically relevant case of having several, though finitely many, tensorial matter fields $A_1, \dots, A_N$ and finitely many tensor fields $G_1, \dots, G_M$ for the underlying geometry, amounting to a matter action 
\begin{equation}
  S_{\textrm{\tiny matter}}[A_1,\dots,A_N;G_1,\dots,G_N) 
\end{equation}
given by a Lagrangian density local in each matter field and ultralocal in each tensor that describes the geometry. All results derived in this article directly generalize to this case of several matter and geometry fields. 

There is, however, one point we wish to draw attention to, in order for the reader to more easily understand this generalization without us actually explicitly performing it: Even in the presence of several matter fields, there is only one principal polynomial $P$ associated with all matter field equations, so that
the all-important triple $(M,G,P)$ extracted from the matter action in the single-fields case generalizes to 
\begin{equation}
   (M, \{G_1, \dots, G_M\}, P)\,,
\end{equation}
in the case of multiple matter fields and geometric fields. More precisely, also in the general case, there
will be just one single principal tensor $P$ in terms of all $G_1, \dots, G_M$ rather than, as one might have erroneously surmised, one such principal tensor for each geometric tensor field. 
One sees this by formally rewriting the system of matter equations that results for several fields $A_1, \dots, A_N$ as one equation for a multiplet $(A_1,\dots,A_N)$ and then calculating the principal polynomial for this overall equation, resulting in one principal polynomial. 
This paper returns to the issue of several fields, both for matter and the geometry, only for the theoretical example in section  \ref{sec_examples}. With these remarks in mind, we return, without loss of generality, 
 to the case $M=N=1$.

\subsection{\new{Massless dispersion relation of test matter field dynamics}}\label{sec_masslessdisp}
We assume that local dynamics for a tensorial matter fields $A$ on a smooth four-dimensional spacetime manifold has been prescribed---motivated by theoretical or phenomenological reasoning---in terms of the action (\ref{intromatter}), which also employs an underived tensor fields $G$ of arbitrary valence such as to produce a scalar density $\mathcal{L}$ of weight one. 

For the purposes of this section, it is simplest to assume that the ensuing matter field equation
\begin{equation}\label{mattersource}
  \frac{\delta S_{\textrm{\tiny matter}}}{\delta A(x)} = 0\,,
\end{equation}
being a tensor density equation of weight one, is linear in the matter field. This assumption corresponds to the requirement we impose for test matter, namely that any solution $A$ of the field equations can be scaled down to $\epsilon A$ by an arbitrarily small factor $\epsilon>0$, so that the source tensor density\begin{equation}
    \frac{\delta S_\textrm{\tiny matter}}{\delta G(x)}\,,
\end{equation}
which will appear on the matter side of the final gravitational field equations, scales down to correspondingly small values. In other words, also the back reaction to the spacetime geometry $G$ can be made arbitrarily small, as it behooves test matter. 

So we obtain test matter field equations (possibly after making implicit information explicit by way of bringing the equations into involutive form \cite{SeilerTucker}) 
\begin{equation}
  Q^{i_1 \dots i_F}_{\mathcal{A}\mathcal{B}}(G(x)) \, (\partial_{i_1} \dots \partial_{i_F} A^{\mathcal{B}})(x)\,\, +  \,\,\textrm{ terms of lower derivative order in } A \,\,=\, 0\,,
\end{equation}
where $\mathcal{A}, \mathcal{B}=1,\dots, R$ are indices labeling a basis of some $R$-dimensional $GL(4)$-representation under which the components of the matter tensor field transform. Note that despite the appearance of only partial derivatives in the highest order term, the left hand side is a tensor density of weight one by construction, with the relevant correction terms being provided by the lower order terms. It follows that $Q^{i_1 \dots i_f}_{\mathcal{A}\mathcal{B}}(G(x))$ is a tensor density of weight one, while the  lower order coefficients, not displayed here, generically are not. 

Any such test matter dynamics provide a dispersion relation for modes of practically infinite frequency, which are physically indistinguishable from massless modes. More precisely, considering a formal Wentzel-Kramers-Brillouin expansion 
\begin{equation}\label{WKB}
  A^{\mathcal{B}}(x) = \textrm{Re}\left\{ \exp(i S(x)/\lambda) \left[a^{\mathcal{B}}(x) + \mathcal{O}(\lambda)\right]\right\}\,,
\end{equation}
one obtains, to lowest order $\lambda^{-F}$ in the approximation, the equation
\begin{equation}\label{WKBeqn}
  Q^{i_1\dots i_F}_{\mathcal{A}\mathcal{B}}(G(x)) k_{i_1} \cdots k_{i_F} a^{\mathcal{B}} = 0\,,
\end{equation}
where the wave covector $k_a(x) := - (\partial_a S)(x)$ is the gradient of the eikonal function $S$. Higher orders in the expansion contain essential information for modes of finite frequency, but equation (\ref{WKB}) precisely captures the behavior in the infinite frequency limit.  

The key question in this limit is for the conditions on $k$ under which there are non-vanishing amplitudes $a^\mathcal{B}$. The answer depends the dimension of any gauge orbits the theory may feature. Indeed, if there is an $s$-dimensional gauge symmetry, which in terms of the lowest order Wentzel-Kramers-Brillouin amplitude $a^\mathcal{A}$ reads
\begin{equation}
  \overline{a}{}^\mathcal{A} = a{}^\mathcal{A} + k_a \chi_{(\sigma)}^{a\mathcal{A}}\,, 
\end{equation} 
for $s$ linearly independent coefficient fields $\chi_{(1)}^{a\mathcal{A}}, \dots, \chi_{(s)}^{a\mathcal{A}}$,
then there is a corresponding $s$-dimensional linear subspace of solutions of (\ref{WKBeqn}) that are pure gauge. 
Using the shorthand $Q_{\mathcal{A}\mathcal{B}}(x,k)$ for the components of the $x$- and $k$-dependent $R\times R$ matrix $Q_{\mathcal{A}\mathcal{B}}^{i_1\dots i_F}(G(x)) k_{i_1} \cdots k_{i_F}$, the condition of having at least one non-vanishing solution $a^\mathcal{A}$ for (\ref{WKBeqn}) that is not purely gauge then amounts to the requirement that  
the adjunct matrix of order $s$, defined by 
\begin{equation}\label{gaugevanish}
Q_{\textrm{adj}}^{[\mathcal{A}_1\dots\mathcal{A}_s][\mathcal{B}_1\dots\mathcal{B}_s]}(x,k) := \frac{\partial^s (\det Q)}{\partial Q_{\mathcal{A}_1\mathcal{B}_1} \cdots \partial Q_{\mathcal{A}_s\mathcal{B}_s}}(x,k)\,,
\end{equation}
must vanish. 
For then the equations of motion have at least $s+1$ linearly independent solutions. But since $s$ of these are pure gauge, this leaves at least one physical solution, which is precisely the condition we wished to impose on the $k$. 
Indeed, the admissible wave covectors are those that satisfy 
\begin{equation}\label{adjunctzero}
  Q_{\textrm{adj}}^{[\mathcal{A}_1\dots\mathcal{A}_s][\mathcal{B}_1\dots\mathcal{B}_s]}(x,k) = 0
\end{equation}
for all $\binom{R}{s}$ independent components of the bilinear map defined by $Q_\textrm{adj}$ on the space of $s$-forms over the $R$-dimensional representation space in which the gauge field takes its values. Each of these independent components is a homogeneous polynomial of degree $(R-s)F$ in the wave covector $k$. At this point, the dispersion relation appears to be given by the condition that a wave covector $k$ be a common root of all these polynomials. Fortunately, however, all of these polynomials share a common factor polynomial density $\widetilde P(k)$, since due to a straightforward generalization of an elegant argument by Itin \cite{Itin}, one has 
\begin{equation}
Q_{\textrm{adj}}^{[\mathcal{A}_1\dots\mathcal{A}_s][\mathcal{B}_1\dots\mathcal{B}_s]}(x,k) = \epsilon^{\sigma_1\dots\sigma_s}\epsilon^{\tau_1\dots\tau_s} 
\chi^{a_1\mathcal{A}_1}_{(\sigma_1)} \cdots \chi^{a_s\mathcal{A}_s}_{(\sigma_s)} \chi^{b_1\mathcal{B}_1}_{(\tau_1)} \cdots\chi^{b_s\mathcal{B}_s}_{(\tau_s)} k_{a_1} \cdots k_{a_s} k_{b_1} \dots k_{b_s} \widetilde P(k)
\end{equation}
for any $s\geq 0$. From the known degree of the homogeneous polynomials that present the components of $Q_\textrm{adj}$, we recognize the common factor polynomial density $\widetilde P$ to be homogeneous of order $FR-(F+2)s$. Most importantly, we see that condition (\ref{adjunctzero}), for the existence of solutions that are not gauge-equivalent to a vanishing solution, is satisfied if and only if 
\begin{equation}
   \widetilde P(x,k) = 0\,,
\end{equation}
which thus emerges as the polynomial dispersion relation for any linear matter theory with gauge-orbits of dimension $s\geq 0$. 

Obviously, we can expand the homogeneous polynomial density as 
\begin{equation}
  \widetilde P(x,k) = \widetilde P^{a_1\dots a_{\deg \widetilde P}}(x) k_{a_1} \cdots k_{a_{\deg \widetilde P}}
\end{equation}
in terms of the components $\widetilde P^{a_1\dots a_{\deg \widetilde P}}(x)$ of \new{a totally symmetric contravariant tensor field density}. Since the principal polynomial is defined, in the first place, only up to a spacetime function factor, we are free to choose an everywhere non-vanishing scalar density $\rho$ of opposite weight in order to obtain the {\it principal tensor field} $P$ with component functions
\begin{equation}\label{rhointro}
  P^{a_1\dots a_{\deg P}}(x) := \rho(x) \, \widetilde P^{a_1\dots a_{\deg P}}(x)\,.
\end{equation}
The choice of $\rho$, however, will only affect what is meant by a massive point particle, see section \ref{sec_massive}, and is to be defined in terms of the tensor field $G$. Since we will finally obtain dynamics for the geometry $G$, also $\rho$ will be determined. 
In any case, $\rho$ has no influence on any field theoretic consideration or massless point particles.

Over the next two subsections, we explain the three matter conditions one must impose in order to start the gravitational closure procedure and show how the kinematical interpretation of the triplet $(M,G,P)$ arises from these.

\subsection{\new{Matter conditions }}\label{subsec_pdr}
The principal tensor field of matter field equations, on which the developments in this paper build, is required to satisfy two hyperbolicity conditions and one energy condition. \new{While these are simply three classical conditions on the classical matter field equations of motion --- more precisely on their principal tensor $P$ and thus indirectly also on the underlying geometry $G$ --- they are found to be indeed necessary in a canonical quantization of the the classical dynamics; see \cite{RS} for a concrete demonstration. In the penultimate  paragraph of the subsection \ref{sec_massive}, we will briefly return to this issue and comment on how classical and quantum considerations together seem to point at precisely the three matter conditions below.} 

\begin{figure}[!h]
\begin{subfigure}{.30\textwidth}
\begin{overpic}[width=4cm]{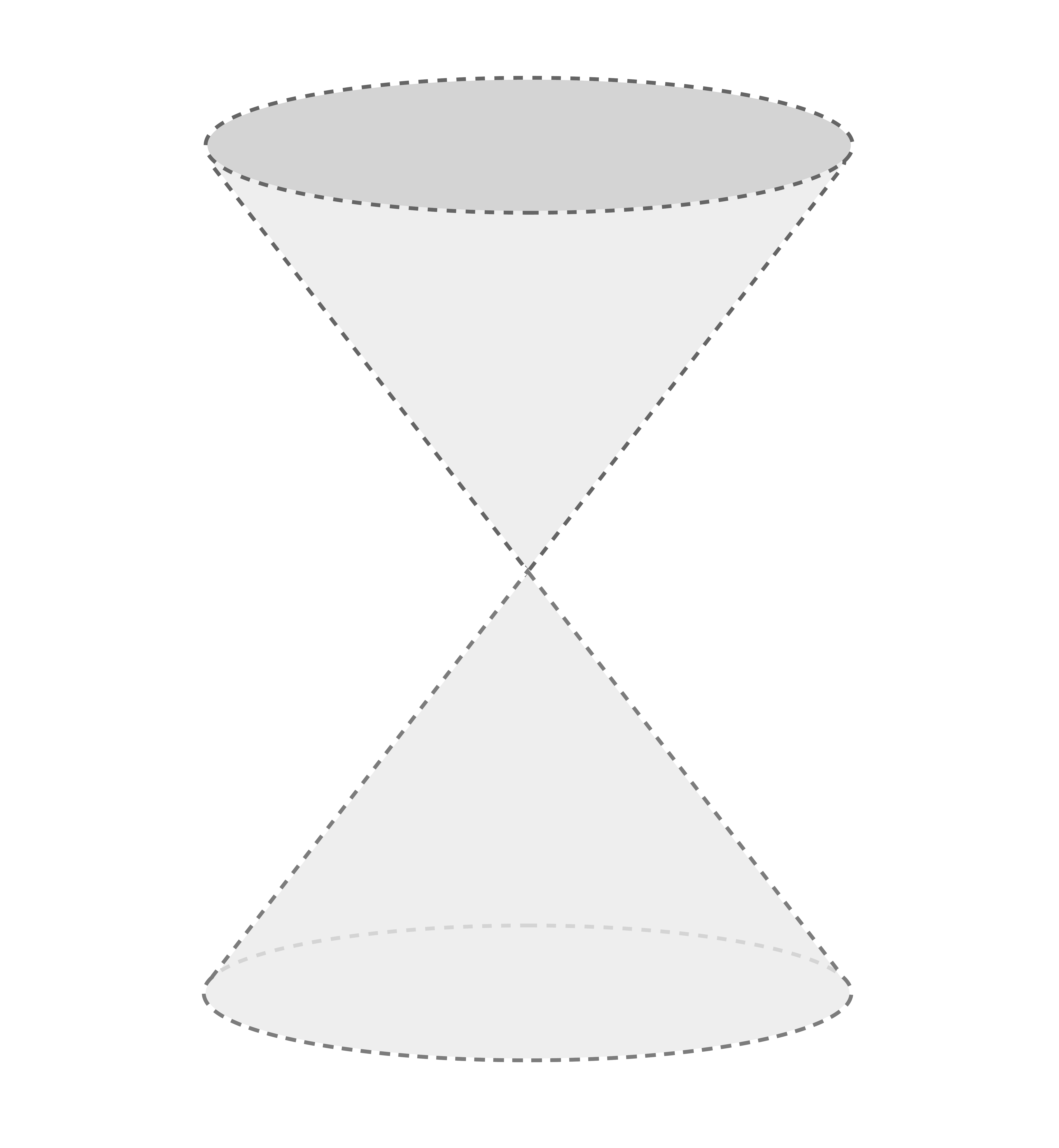}
  \put(100,105) {$C_1$}
  \put(100,15) {$C_2$}
\end{overpic}
\caption{The two hyperbolicity cones of a hyperbolic second degree principal polynomial\label{subfig_metrichypcones}}
\end{subfigure}
\begin{subfigure}{.30\textwidth}
\begin{overpic}[width=4cm]{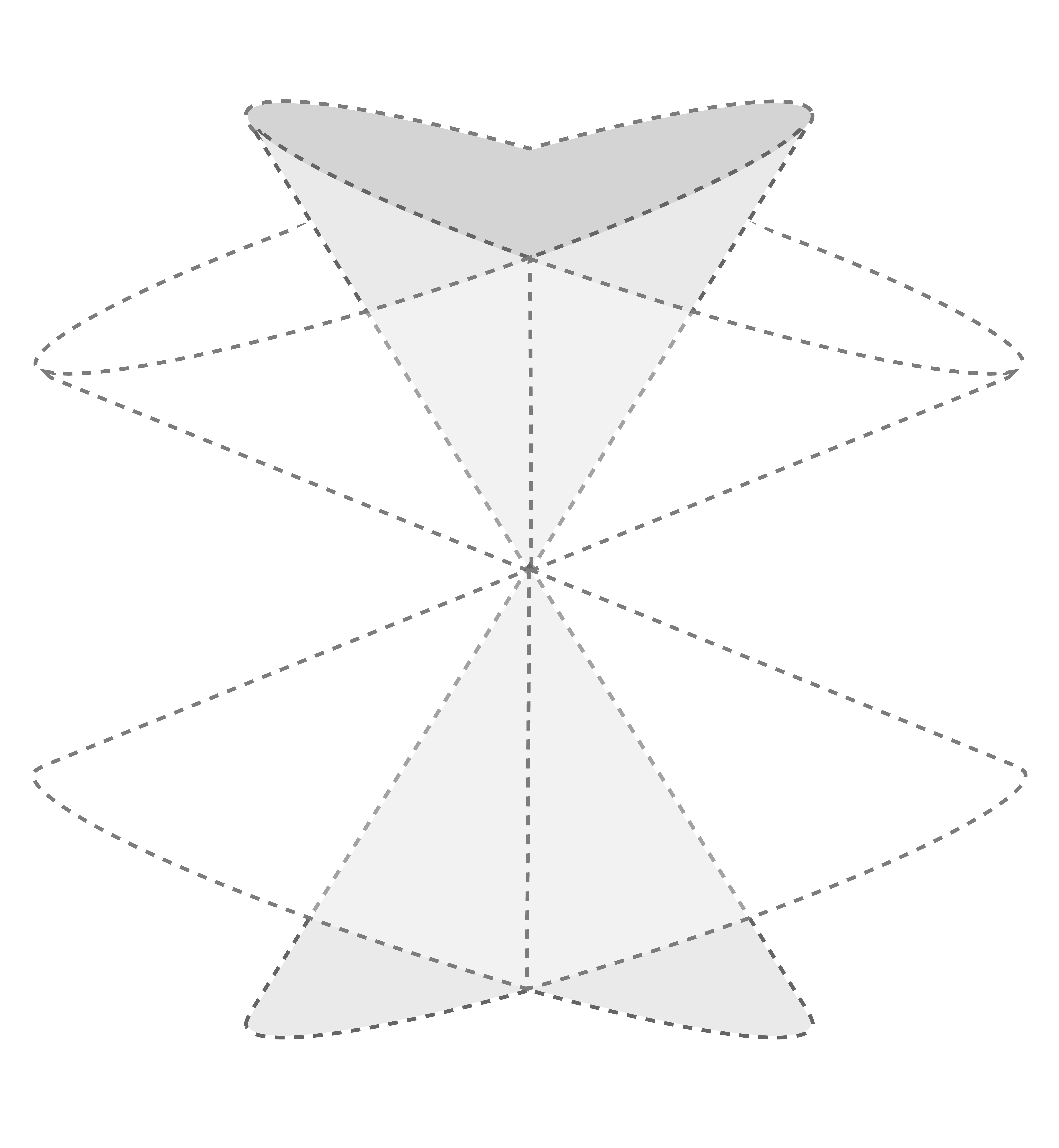}
   \put(95,105) {$C_1$}
  \put(95,15) {$C_2$}
\end{overpic}
\caption{Hyperbolicity cones of a hyperbolic reducible fourth degree principal polynomial\label{subfig_bimetrichypcones}}
\end{subfigure}
\begin{subfigure}{.36\textwidth}
\includegraphics[width=4cm]{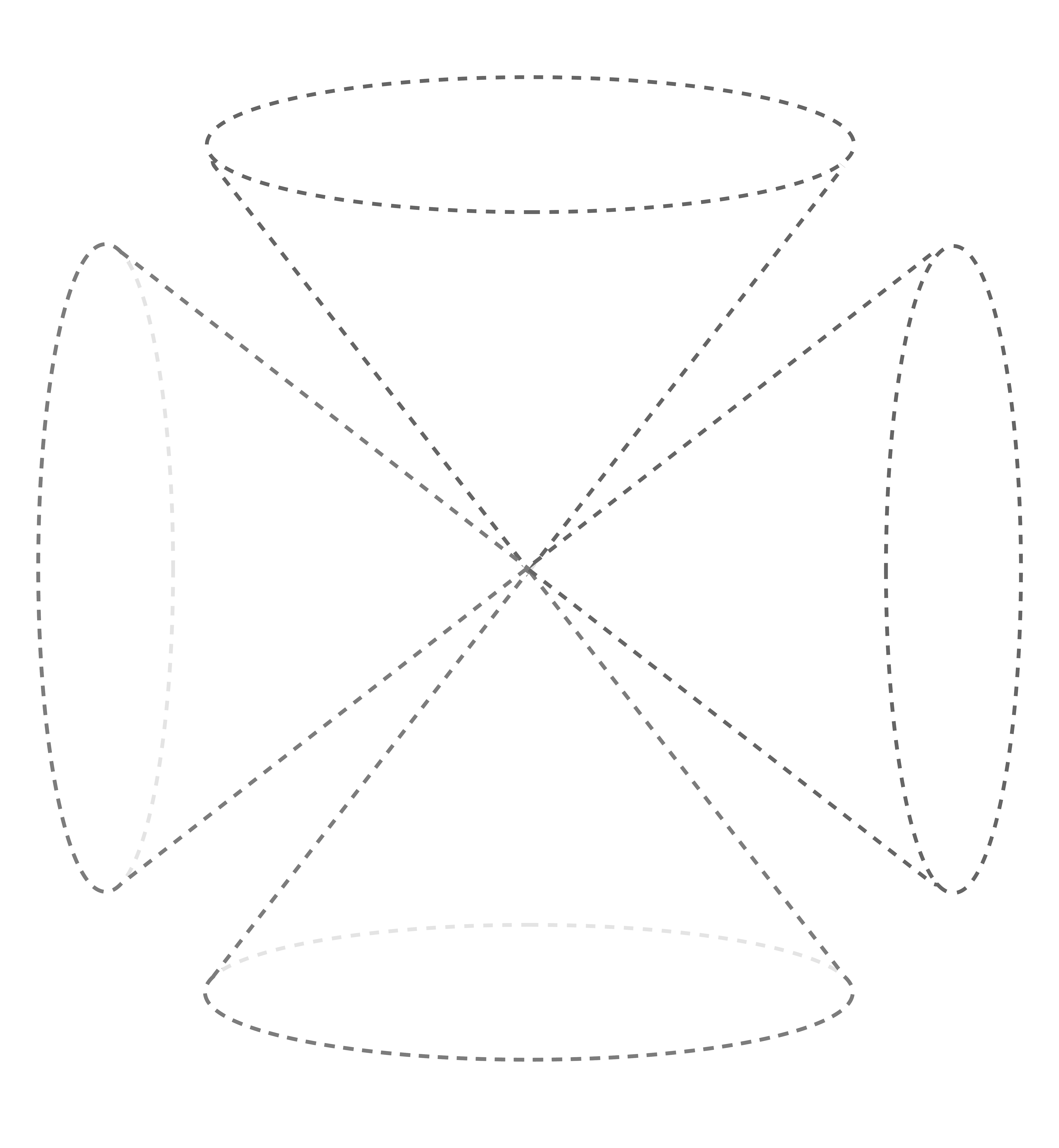}
\caption{A non-hyperbolic fourth degree principal polynomial, obviously featuring no hyperbolicity cones\label{subfig_nohypcones}}
\end{subfigure}
\caption{Hyperbolicity cones of various polynomials\label{fig_hypcones}}
\end{figure}

\begin{center}{\it Matter condition 1: Predictivity}\end{center} 
The first technical condition is the {\it hyperbolicity of the principal polynomial} $P(x)$ at every point $x\in M$, which is directly enforced by the physical assumption that there be an initial value formulation of the classical field equations \cite{SeilerBook}. The polynomial $P(x)$ is called hyperbolic if there exists a covector $h\in T^*_pM$ such that $P(x)(h)\neq 0$ and the equation $P(x)(q+\lambda h)=0$ has only real solutions $\lambda$ for any further covector $q\in T^*_xM$. But as soon as one such covector $h$ exists, there is always an open and convex cone $C_x(P,h)$ that contains all hyperbolic covectors that lie together with $h$ in one connected set \cite{Garding}. But if indeed there is any such non-empty hyperbolicity cone, and thus the polynomial $P(x)$ hyperbolic, then there is always an even number of distinct hyperbolicity cones, see figures \ref{subfig_metrichypcones} and \ref{subfig_bimetrichypcones}.


It is easy to see that if the polynomial $P(x)$ is reducible, meaning that it can be written as a finite product 
\begin{equation}\label{prod3}
   P(x) = P_1(x) P_2(x) \cdots P_f(x)
\end{equation}
of lower degree polynomials, then $P(x)$ is hyperbolic if and only if each of the lower degree polynomials is hyperbolic, and that the various hyperbolicity cones of $P(x)$ are obtained by the intersections 
\begin{equation}\label{prod4}
  C(P,h) = C(P_1,h) \cap \dots \cap C(P_f,h) 
\end{equation}
of the various hyperbolicity cones of the lower degree polynomials, compare figures \ref{subfig_bimetrichypcones} and \ref{subfig_nohypcones} for examples. Clearly, $C(P,h)=\emptyset$ unless $h$ is a hyperbolic covector of every factor polynomial. There is obviously no loss of information incurred by removing repeated polynomial factors if such happen to occur, but doing so is indeed technically important \cite{RRS} for the formulation of our second condition on the polynomials $P(x)$. In the following, we will therefore assume that repeated factors have been removed from $P$.

\begin{figure}[!h]
\begin{subfigure}{.49\textwidth}
\begin{overpic}[width=5cm]{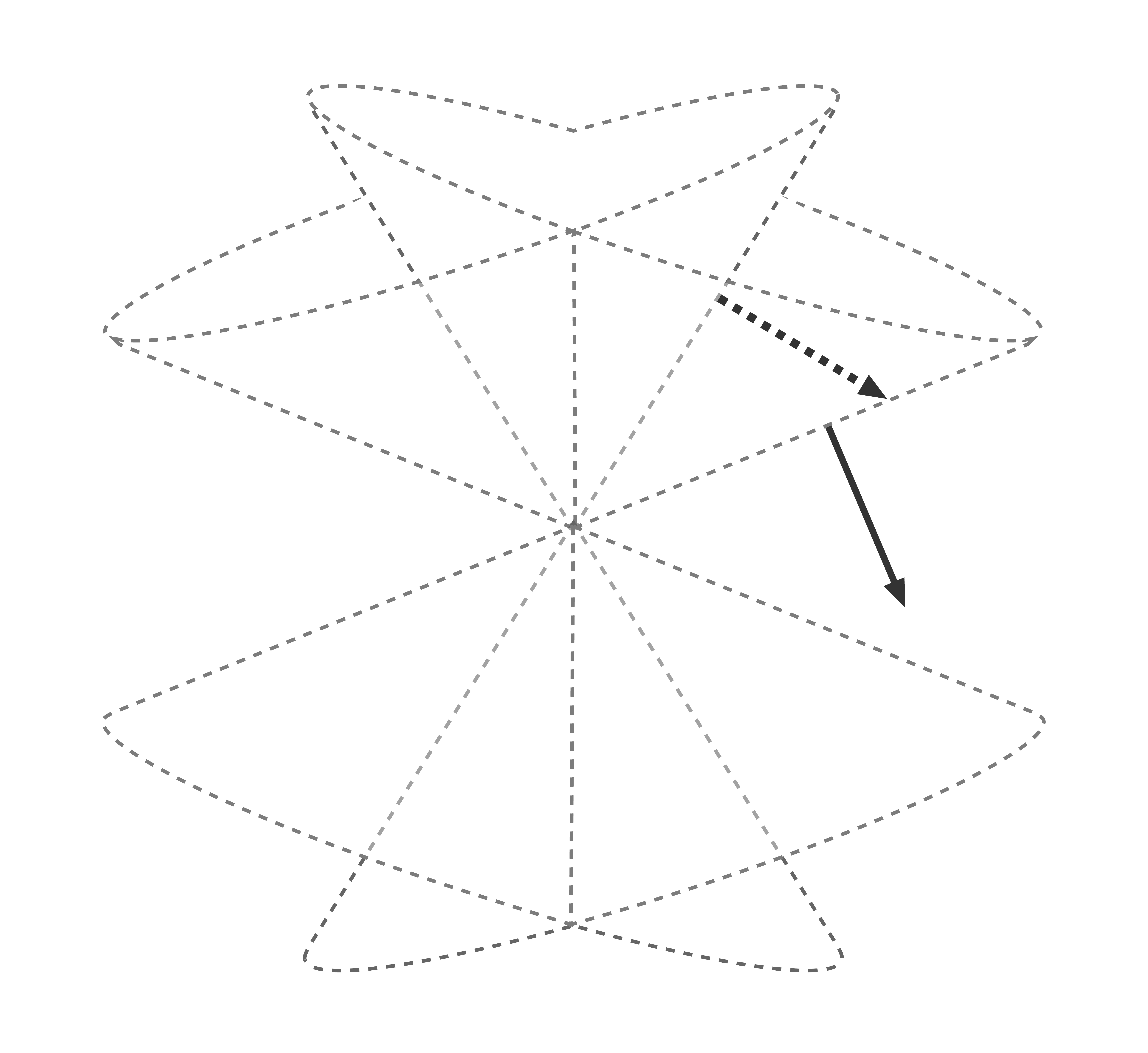}
  \put(110,110) {}
\end{overpic}
\caption{Null surface of a hyperbolic reducible principal polynomial $P$ in cotangent space;\\ with typical gradient (co-co-)vectors\label{subfig_leftGauss}}
\end{subfigure}
\begin{subfigure}{.49\textwidth}
\includegraphics[width=5cm]{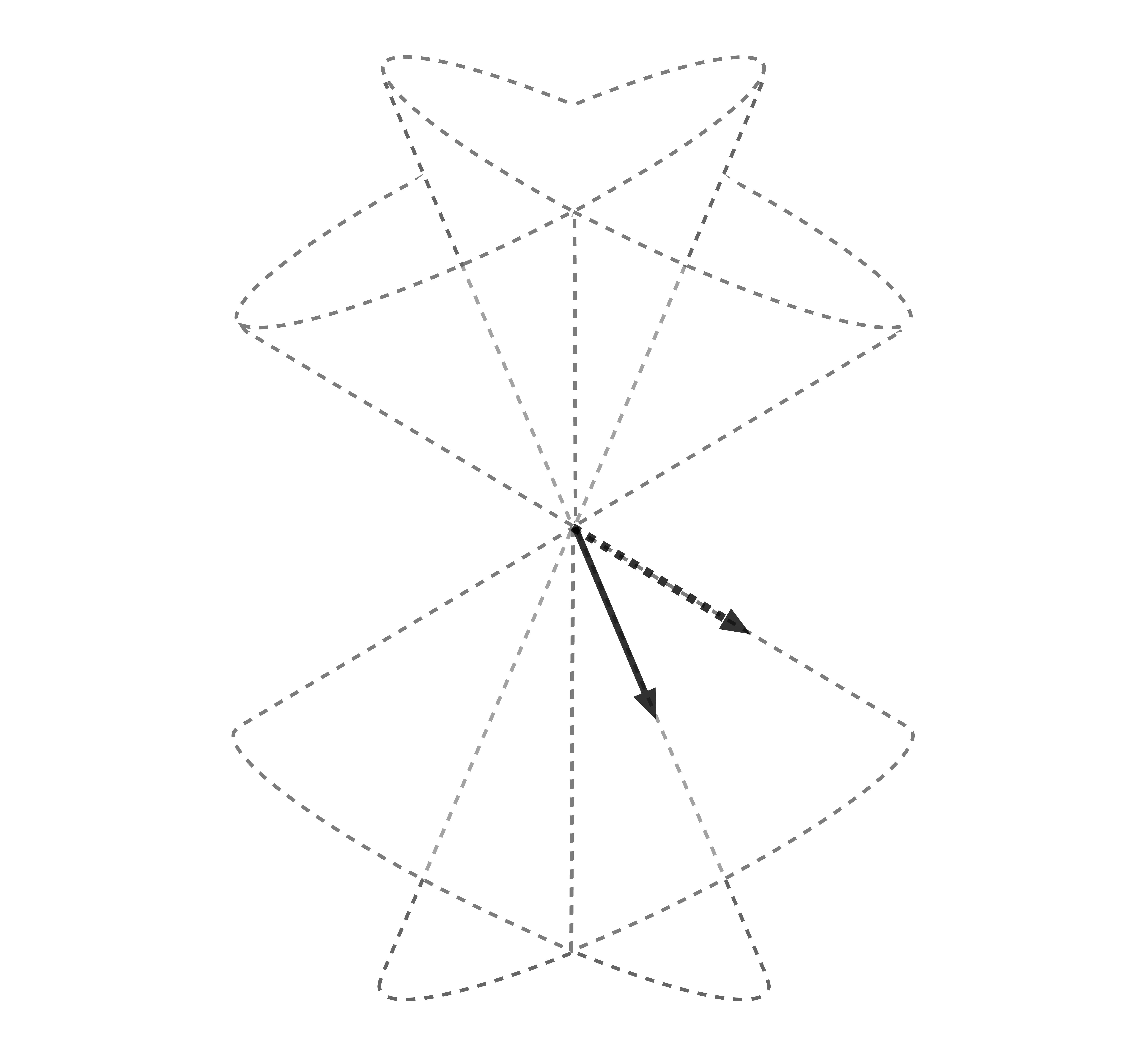}
\caption{Null surface of the dual polynomial $P^\#$ in tangent space; containing, by definition, the gradient vectors to the $P$-null surface\label{subfig_rightGauss}}
\end{subfigure}
\caption{Gauss map sending $P$-null covectors to $P^\#$-null vectors.\label{fig_Gauss}}
\end{figure}

\begin{center}{\it Matter condition 2: Momentum-velocity duality}\end{center} 
The second technical condition is the {\it hyperbolicity of the dual polynomial}
\begin{equation}
P^\#(x): T_xM \longrightarrow \mathbb{R}\,,\qquad P^\#(x) := P_1^\#(x) P_2^\#(x) \cdots P_f^\#(x)\,,
\end{equation}
where the $P_1^\#(x), \dots, P_f^\#(x)$ are polynomial maps $T_xM \longrightarrow \mathbb{R}$ of minimal degree such that for all $k$ in the set
\begin{equation}
   N^\textrm{smooth}_i(x) := \{k\in T_x^*M \,|\, P_i(x,k)=0 \textrm{ and } \frac{\partial P_i}{\partial k}(x,k) \neq 0\}\,,
\end{equation}
the
precisely the gradients $\partial P_i/\partial k \in T_xM$ (shown in figure \ref{subfig_leftGauss} as gradients to the null surfaces in cotangent space and in figure \ref{subfig_rightGauss} as elements of the tangent space) are the roots of $P_i^\#$:
\begin{equation}
  P_i^\#(x,\frac{\partial P_i}{\partial k}(x,k)) = 0\qquad\textrm{ for all} \quad k\in N^\textrm{smooth}_i(x).  
\end{equation}
While the polynomial $P^\#(x)$ is thus only determined up to a real factor function, its roots, and thus its hyperbolicity, are unaffected by this ambiguity and thus well-defined. It is shown in \cite{RRS} that the existence of a dual $P^\#(x)$ hinges on the hyperbolicity of $P(x)$, which is however guaranteed by the first matter condition above. 

Technically, the now additionally required hyperbolicity of $P^\#(x)$ can thus be understood as a sufficient condition for $P(x)$ to be recoverable from $P^\#$ as the double dual, such that at each $x\in M$ we have the proportionality 
\begin{equation}
  P(x) \sim P^{\#\#}(x)\,.
\end{equation}
The physical meaning of the dual polynomial becomes apparent by noting that the characteristic curves $x: \mathbb{R} \to M$ of the initial matter field equations are, by definition, stationary with respect to the Hamiltonian action
\begin{equation}
  S_\textrm{\tiny massless}[x,k,\rho] := \int \mathrm d\lambda \big[k_a(\lambda) \dot x^a(\lambda) - \rho(\lambda) P(x(\lambda),k(\lambda))\big]\,,
\end{equation}
which has been shown \cite{RRS} to be equivalent to the Lagrangian action
\begin{equation}
  S_\textrm{\tiny massless}[x,\mu] := \int \mathrm d\lambda \, \mu(\lambda) P^\#(x(\lambda),\dot x(\lambda))
\end{equation}
for any hyperbolic $P$. Hyperbolicity of both $P(x)$ and $P^\#(x)$ thus ensures the free passage back and forth between the Hamiltonian to the Lagrangian formulation in case of characteristic curves, which physically correspond to the trajectories of massless particles. In other words, the said bihyperbolicity ensures that, up to scale, there is a momentum associated with each massless particle velocity, and vice versa.

\begin{figure}[!h]
\begin{subfigure}{.49\textwidth}
\begin{overpic}[width=7cm]{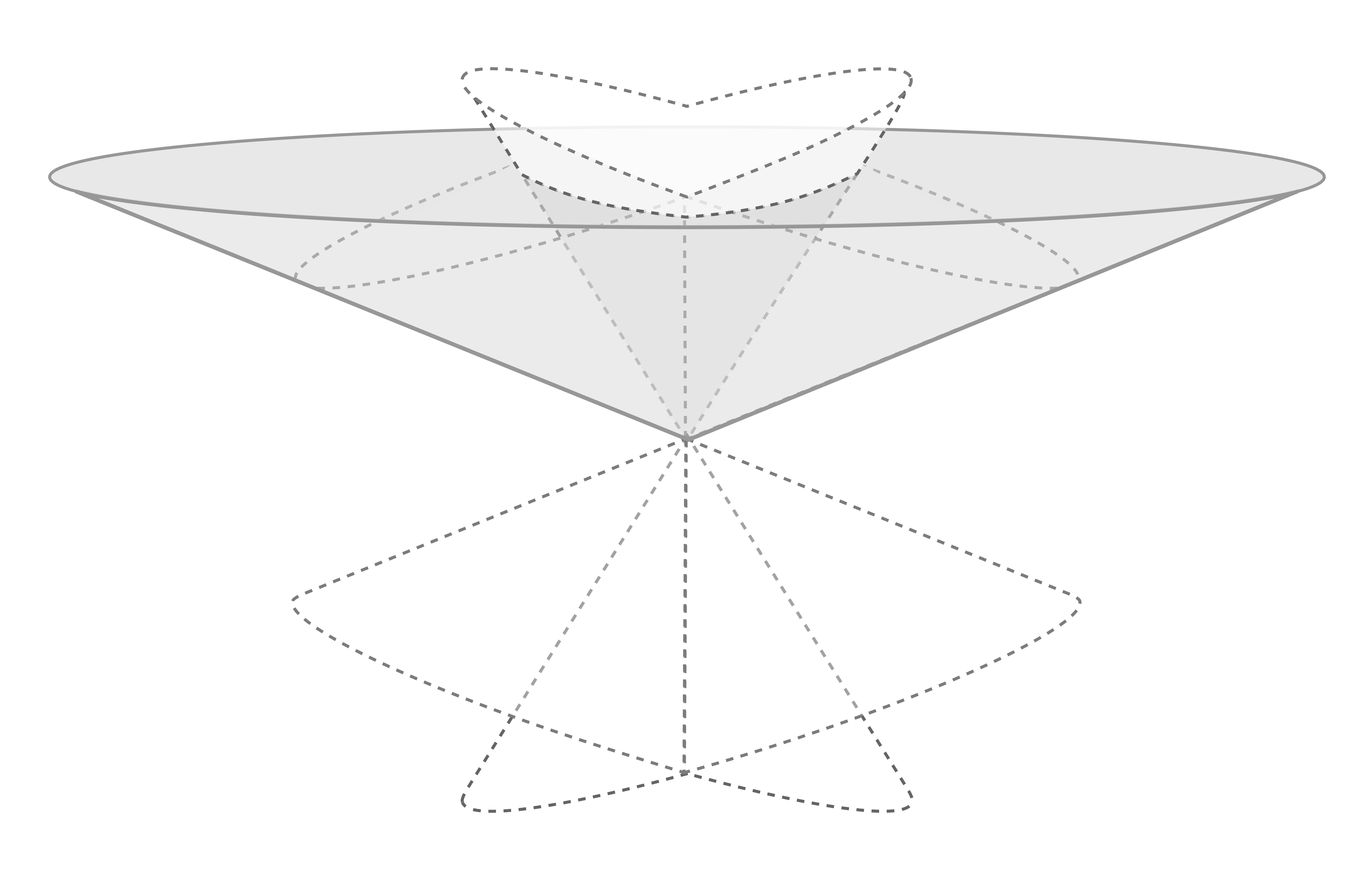}
  \put(190,110) {$\mathcal{O}_x^+$}
\end{overpic}
\caption{Cone covering all momenta of positive energy as unanimously judged by all observers
\label{subfig_positivecone}}
\end{subfigure}
\begin{subfigure}{.49\textwidth}
\begin{overpic}[width=5cm]{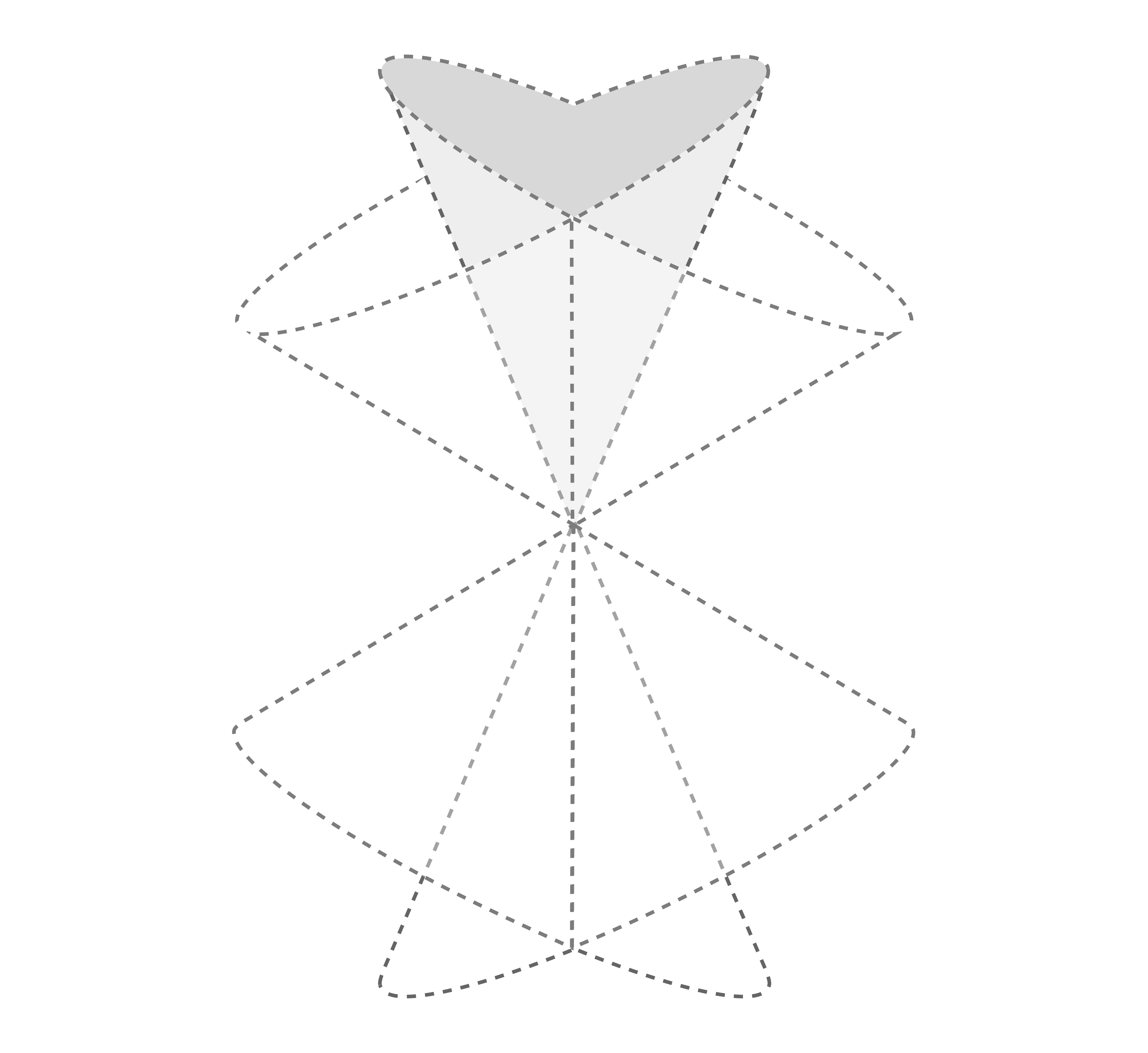}
  \put(100,113) {$\mathcal{O}_x$}
\end{overpic}
\caption{Cone containing all tangent vectors to observer worldlines through one point\label{subfig_observercone}}
\end{subfigure}
\caption{Positive energy cone $\mathcal{O}_x^+$ as the dual of the observer cone $\mathcal{O}_x$ \label{fig_energyandobservers}}
\end{figure}

\begin{center}{\it Matter condition 3: Energy distinction.}\end{center}
Having established the physical reasoning behind the condition of hyperbolicity for both the cotangent-space polynomials $P(x)$ and the tangent-space polynomials $P^\#(x)$, we are now prepared to turn to the physical meaning one must attach to the respective hyperbolicity cones. To this end, we recall the insight quoted above, namely that the momenta $k$ of massless particles satisfy the dispersion relation
\begin{equation}\label{masslessPk}
P(x,k) = 0\,.
\end{equation}
In order to divide the set of all such massless momenta $k$, in an observer-independent way, into momenta of either positive or negative energy, we now wish to find the largest possible set of local observers that can still agree on such a division. More precisely, we wish to find the largest possible open set $O_x$ in each tangent space $T_xM$ of the spacetime manifold that contains all tangent vectors $U$ to observer worldlines such that for any particular nonvanishing massless momentum $k$, 
\begin{equation}
  \textrm{either} \qquad k \in \mathcal{O}_x^+ \qquad \textrm{or} \qquad k \in -\mathcal{O}_x^+\,,
\end{equation}
where the closed dual cone
\begin{equation}
  \mathcal{O}_x^+ := \left\{q \in T_x^*M \,|\, U(k)>0 \textrm{ for all } U \in \mathcal{O}_x\right\}\,,
\end{equation}
see figure \ref{subfig_positivecone} for an illustration, 
implements the said observer-dependent positive energy condition when intersected with the set of all nonvanishing massless momenta. 
Formally, we require that the cone $N_x$ of massless momenta at every spacetime point $x$ decomposes into disjoint pieces
\begin{equation}\label{distinction}
  N_x\backslash\{0\} \,\,=\,\, N_x^+ \,\,\dot\cup \,\, N_x^-\,,
\end{equation}
where $N^\pm_x := N_x \cap (\pm\mathcal{O}_x^+)$. So what is the largest cone $\mathcal{O}_x$ on can choose? 

If the above conditions can be satisfied at all, it turns out that any one of the hyperbolicity cones of $P^\#(x)$, see figure \ref{subfig_observercone},  provides the required largest set one can choose in order to satisfy the {\it energy condition} (\ref{distinction}). This only leaves us with a choice between the finitely many hyperbolicity cones of $P^\#(x)$ at each point $x$ of the spacetime manifold. A smooth choice throughout the manifold is clearly provided by choice of a smooth vector field $T$ that is everywhere hyperbolic with respect to $P^\#$, such that we obtain a smooth distribution of
\begin{equation}
\textrm{future-directed observer cones } \quad O_x = C_x(P^\#,T)\,.
\end{equation}

Note that all three matter conditions above only employ the roots of the principal polynomial $P$ and its dual $P^\#$ at each spacetime point. Indeed, even the observer cones $O_x$ are defined entirely in terms of the roots of  the dual polynomial, although all tangent vectors they contain are non-roots. In the following subsection, we will now complete the kinematics by, first, defining the kinematics of massive particles within the above framework and, second, by employing a thus emerging Legendre map in order to define local observer frames.

\subsection{\new{Massive dispersion relation and local observer frames}}\label{sec_massive}
\begin{figure}[h]
\begin{subfigure}{.49\textwidth}
\includegraphics[width=3.6cm]{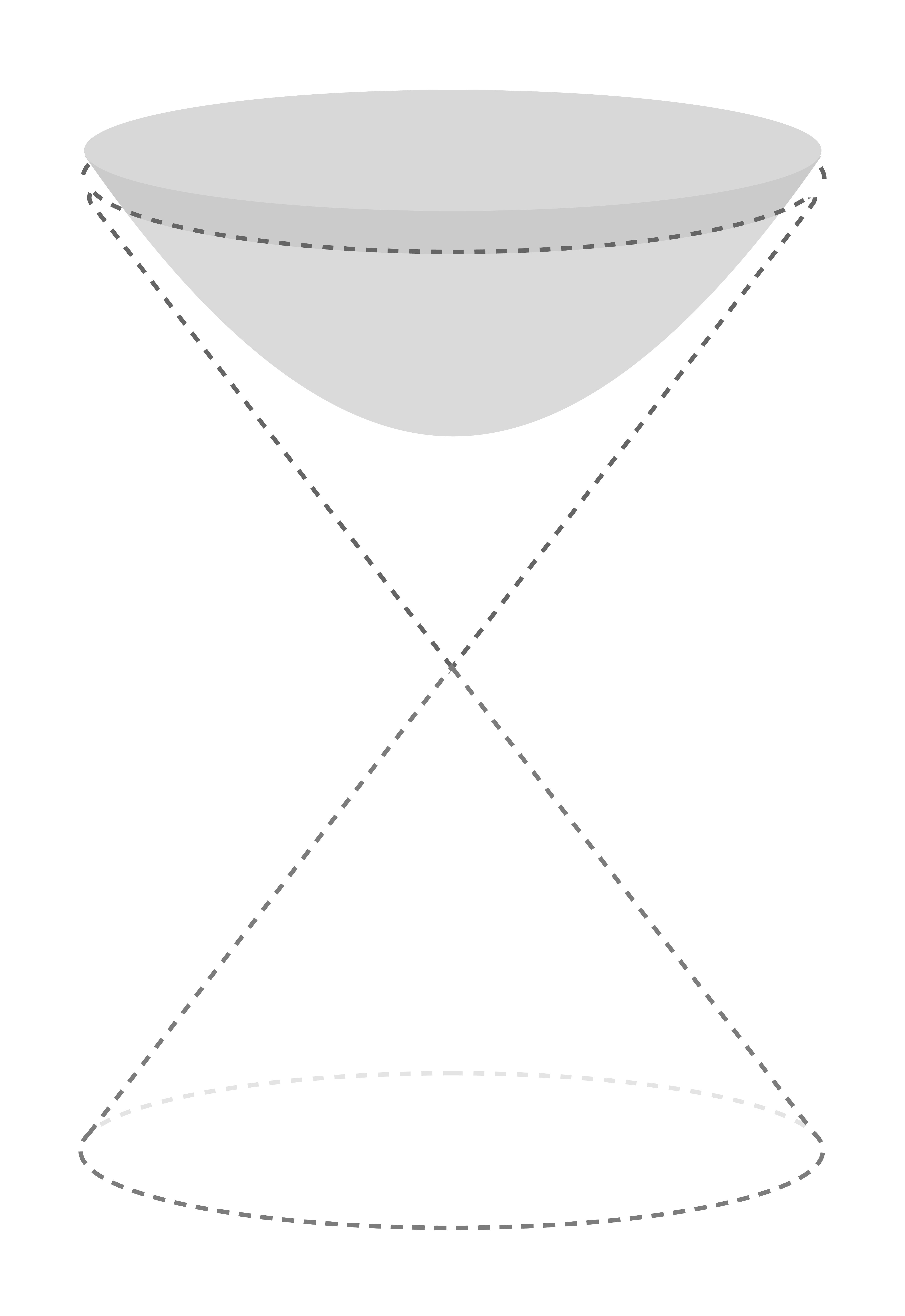}
\caption{Quadric mass shell of a second degree principal polynomial $P_x$ satisfying\\
the three matter conditions\label{subfig_metricmassshell}}
\end{subfigure}
\begin{subfigure}{.49\textwidth}
\includegraphics[width=5cm]{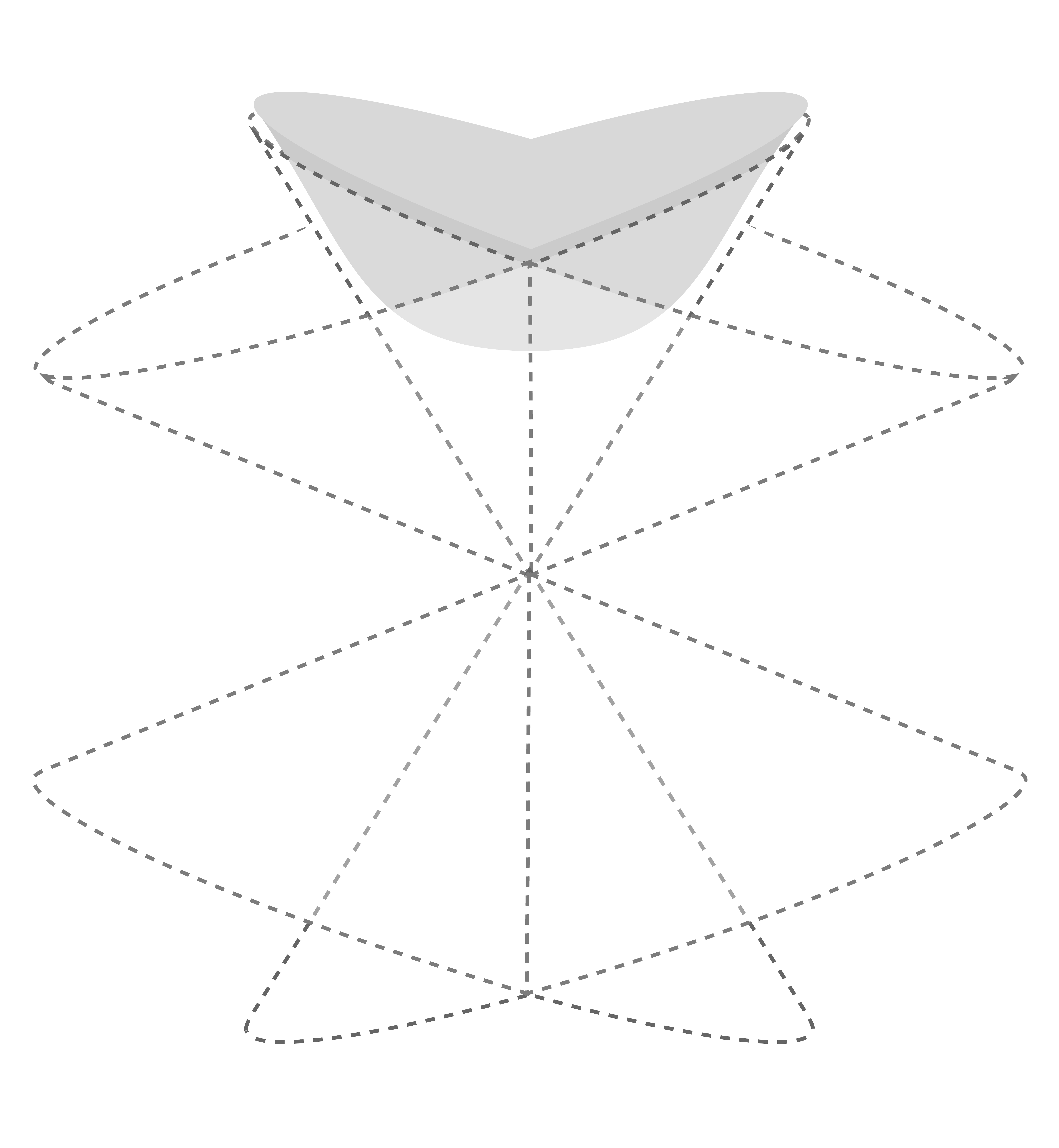}
\caption{Quartic mass shell of a fourth degree principal polynomial $P_x$ satisfying\\ the three matter conditions\label{subfig_areametricmassshell}}
\end{subfigure}
\caption{Examples of positive energy mass shells\label{fig_massshells}}
\end{figure}

To the smooth choice of observer cones $O_x$ on each tangent space, for a principal polynomial that satisfies all three matter conditions imposed in the previous subsection, corresponds a smooth choice of a hyperbolicity cone in cotangent space, the so-called cone $C_x$ of positive energy massive momenta \cite{RRS} that satisfies
\begin{equation}
  C_x \subseteq O_x^\perp\,.
\end{equation}
It is a general result \cite{Garding} that hyperbolicity cones are open convex cones, whose boundary is null with respect to the defining polynomial while the interior has constant sign. Since so far we have only employed the roots of the principal polynomial, we can freely scale it by a sign such as to conventionally achieve
\begin{equation}
  P(x,C_x)>0\qquad\textrm{ for all } x\in M\,. 
\end{equation}
While the hyperbolicity condition on $P$ generalizes the Lorentzian signature condition for an inverse metric, and the hyperbolicity condition on the dual $P^\#$ that of a metric itself, it is the above sign convention that generalizes the mainly-minus signature convention of Lorentzian geometry. With this choice made, we now define the mass $m>0$ of a momentum $q \in C_x$ by
\begin{equation}
   P(x,q) = m^{\deg P}\,,
\end{equation}
see figure \ref{fig_massshells} for the genesis of quadric and quartic mass shells.
Note that this definition of point particle rest mass is the first definition that depends on the choice of the scalar density $\rho_G$ in (\ref{rhointro}), which converts the tensor field density $\widetilde P$ into a tensor field $P$. 

As in the massless case, this dispersion relation is easily implemented as a constraint in the Hamiltonian action \cite{RRS}
\begin{equation}
  S_\textrm{\tiny massive}[x,q,\rho] := \int \mathrm d\lambda\left[q_a(\lambda) \dot x^a(\lambda) - \rho(\lambda) \ln P(x(\lambda),m^{-1} q(\lambda))\right]\,,
\end{equation}
from which the momentum $q$ can be eliminated---by way of an injective Legendre map 
\begin{equation}\label{Legendremap}
\ell_x: C_x \longrightarrow T_xM\,,\quad \ell_x(q) := \frac{1}{\deg P} \frac{\partial \ln P}{\partial q}(x,q)
\end{equation}
at each point $x$ of the spacetime manifold $M$, for which the inverse $\ell^{-1}_x: \ell_x(C_x) \longrightarrow C_x$ is guaranteed to exist due to the three matter conditions imposed in the previous subsection---such that the Lagrangian action for the trajectory $x$ of a positive energy particle of mass $m$ emerges \cite{RRS} as
\begin{equation}\label{massivepointparticle}
   S_\textrm{\tiny massive}[x] := \int \mathrm d\lambda\, m \sqrt[\deg P]{P^*\left(x(\lambda),\dot x(\lambda)\right)}
 \end{equation} 
 in terms of the decidedly non-polynomial map
 \begin{equation}
 P^*(x): \ell_x(C_x) \longrightarrow \mathbb{R}\,,\qquad P^*(x,v) := P\left(x,\ell^{-1}_{x}(v)\right)^{-1}\,.
\end{equation}
Note that the massive point particle action (\ref{massivepointparticle}) is invariant under strictly monotonously increasing reparametrizations of the trajectory. This observation also affords us the final piece of information required for the construction of {\it local observer frames}, namely the observation that parametrizations with
\begin{equation}
  P^*(x(\lambda),\dot x(\lambda)) = 1
\end{equation}
are distinguished because they correspond to massive particle momenta 
\begin{equation}
  q(\lambda) = m \dot x(\lambda) 
\end{equation}
proportional to the particle velocity, with the proportionality given by the particle rest mass. Employing such parameters as the definition of proper time, we define an observer worldline $x$ by the requirements
\begin{equation}
  \dot x(\lambda) \in O_{x(\lambda)} \qquad \textrm{ and } \qquad P^*(x(\lambda),\dot x(\lambda)) = 1\,, 
\end{equation}
and identify the purely spatial directions $S(\lambda)\subset T_{x(\lambda)}M$ seen by an observer at $x(\lambda)$ by
\begin{equation}
  \ell_{x(\lambda)}(\dot x(\lambda))(S(\lambda)) = 0\,.
\end{equation}

\new{Note that the above constructions show that the three physicality conditions of section \ref{subsec_pdr} are {\it sufficient} for the formulation of observer frames that are compatible with the causality of the original matter field equations. Together with their being {\it necessary} for the canonical quantizability of the same matter field dynamics, this provides one possible circumscription of their physical motivation.}

In our below derivation of the gravitational closure equations --- which take a test matter action satisfying the three matter conditions as input and yields the underlying gravitational dynamics as output --- the key information contained in the matter dynamics trickles down to the gravitational side exclusively through the Legendre maps $\ell_x$, whence (\ref{Legendremap}) presents the most important result of this review section from a practical point of view. 

\section{Canonical geometry}\label{sec_canonicalkinematics} 
\new{In this section, we employ the kinematics implied by given matter field dynamics in order to foliate the spacetime manifold into initial data surfaces. Calculating the commutation relations between normal and tangential deformation operators acting on functionals of initial data hypersurface embedding maps, we obtain the hypersurface deformation algebra for any spacetime structure $(M,G,P)$ that satisfies the matter conditions in \ref{subsec_pdr}. We then devise an associated bundle to the frame bundle of the manifold that serves to parametrize a canonical geometry that mimics all possible projections of the spacetime geometry to the leaves of the foliation. 
In contrast to previous work, where possibly non-linear constraints on the canonical geometry had been left as almost intractable subsidiary conditions in the solution of the gravitational closure equations, our associated bundle technique now allows to capture these constraints automatically. This is the conceptual and technical basis for the construction of the canonical phase space for the geometry, at the beginning of the next chapter and throughout the remainder of this paper.}

\subsection{Spacetime foliation and induced geometry}   \label{subsec_transition}
Foliating the spacetime into leaves of initial data hypersurfaces and inducing a canonical geometry, a standard technique in general relativity, straightforwardly extends to manifolds $(M,G,P)$ whose structure arises from canonically quantizable matter field actions. In order to fix the notation and to devise a way to project spacetime geometries $G$ of arbitrary valence to initial data surfaces, we quickly collect the relevant constructions.

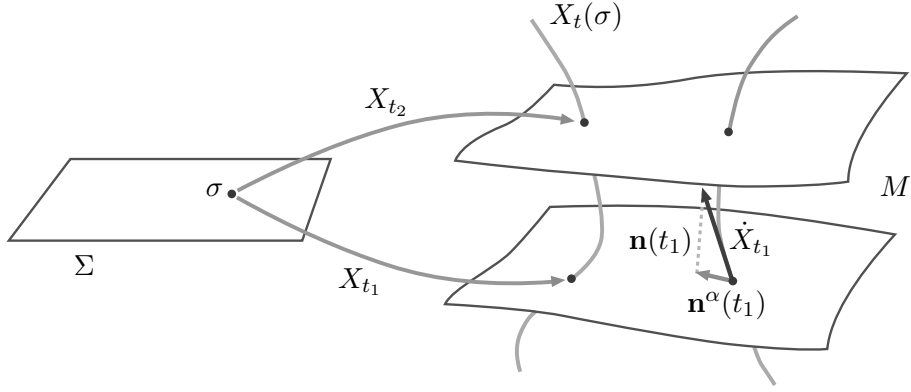
\begin{figure} 
\centering
\begin{overpic}[width=13cm]{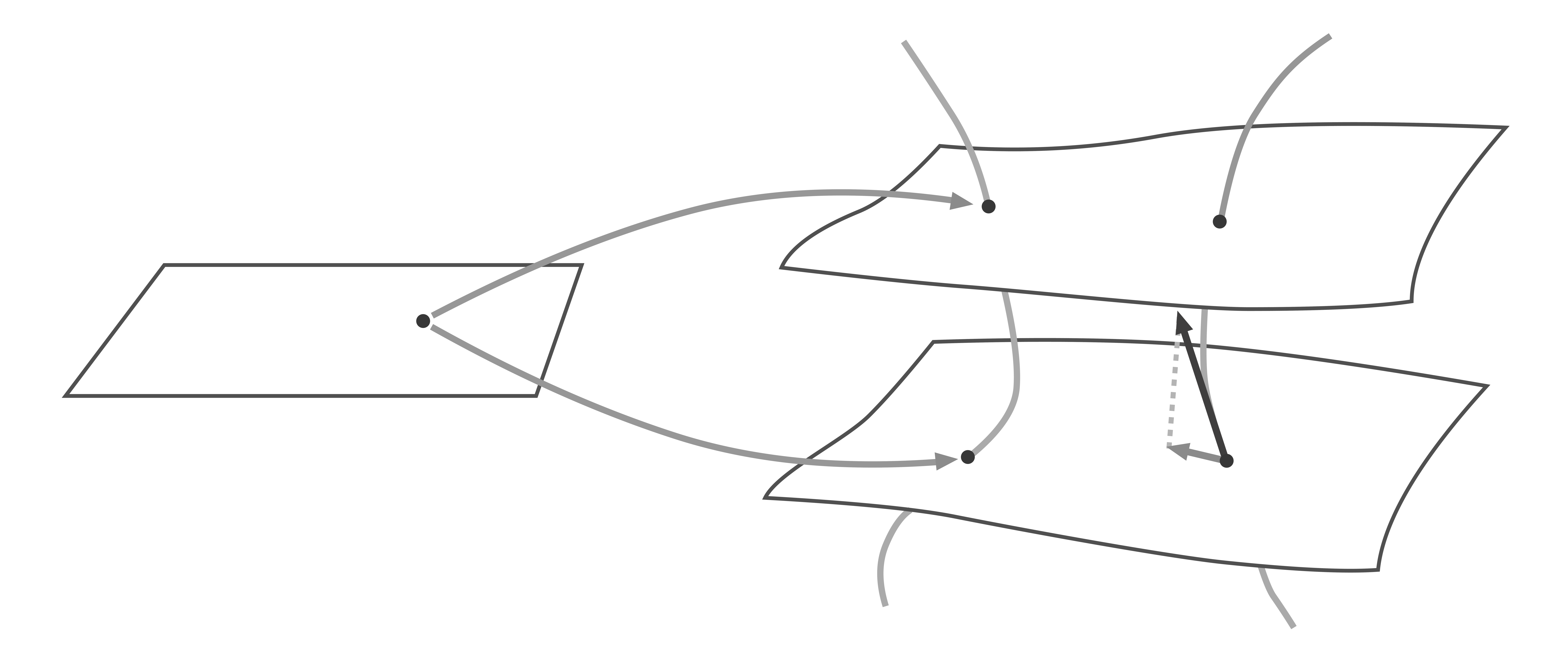}
    \put(90,80) {$\sigma$}
    \put(220,145) {$X_t(\sigma)$}
	\put(345,80) {$M$}
	\put(40,50) {$\Sigma$}
    \put(140,45) {$X_{t_1}$}
    \put(150,113) {$X_{t_2}$}
    \put(288,60){$\dot{X}_{t_1}$}
    \put(250,60) {$\mathbf{n}(t_1)$}
    \put(272,36) {$\mathbf{n}^\alpha(t_1)$}
\end{overpic}
\caption{Embedding of the immutable three-dimensional screen manifold $\Sigma$ by a family of embedding maps $X_t$ into a smooth spacetime manifold $M$ of appropriate topology, yielding the leaves of a foliation of that spacetime into initial data hypersurfaces $X_t(\Sigma)$ \label{fig_embedding}}
\end{figure}

Let $X_t:  \Sigma \hookrightarrow M$ be a one-real-parameter family of maps embedding a three-dimensional manifold $\Sigma$ such that $M$ is foliated into hypersurfaces $X_t(\Sigma)$ with everywhere hyperbolic conormal $\epsilon^0(t,\sigma)$ for $\sigma\in\Sigma$. Employing coordinates $y^\alpha$ on $\Sigma$, 
we define the one-parameter families of spacetime vectors
\begin{equation}\label{ezero}
e_0(t,\sigma) := \ell_{X_t(\sigma)}(\epsilon^0(t,\sigma))\qquad \textrm{and}\qquad e_{\alpha}(t,\sigma) := X_{t\,*}((\partial/\partial y^\alpha)_\sigma)
\end{equation} 
for $t\in\mathbb{R}$ and $\sigma\in\Sigma$, see figure \ref{fig_embedding}.
With the normalization condition $P(X_t(\sigma),\epsilon^0(t,\sigma))=1$, these provide the so-called orthogonal projection frame field along each embedded hypersurface $X_t(\Sigma)$. 
The frames $e_0(t,\sigma),\dots,e_3(t,\sigma)$, together with their unique dual frames $\epsilon^0(t,\sigma),\dots,\epsilon^3(t,\sigma)$, allow to project  spacetime tensors of arbitrary valence to the manifold $\Sigma$. 

In the context of this article, we will perform such projections for the spacetime tangent vector field $\dot X_t$ constructed from the family of embedding maps, the spacetime tensor field $G$ and the principal tensor $P$. We will discuss these, in turn, below. The manifold $\Sigma$ thus becomes a kind of three-dimensional cinema screen on which the evolution of the four-dimensional spacetime geometry is shown as a movie in the foliation parameter $t$. Note that the entire construction is conceptually standard, but that the Legendre maps $\ell_x$  generically are non-linear for the spacetime geometries $(M,G,P)$ we consider. 

Now more precisely, consider first the vector field $\dot X_t$, which is the tangent vector field to the congruence of spacetime curves that correspond to points that do not move on the manifold $\Sigma$ as the foliation parameter increases. Its projection to $\Sigma$, see figure \ref{fig_embedding}, gives rise to two one-parameter families of fields, namely the induced lapse and shift fields
\begin{equation}
\mathbf{n}(t) := \epsilon^0(t)(\dot X_t) \qquad\textrm{and}\qquad \mathbf{n}^\alpha(t) := \epsilon^\alpha(t)(\dot X_t)\,.
\end{equation}

Secondly, we perform the projection of the spacetime geometry $G$ to several one-parameter families of tensors on $\Sigma$, which is an important intermediate step towards setting up the gravitational closure  equations for any $(M,G,P)$. Their components are practically obtained \cite{SW2} by inserting either the frame field $e_0(t\,,\sigma)$ or $e_\alpha(t,\sigma)$ into a slot of $G$ that requires a vector, and correspondingly either $\epsilon^0(t,\sigma)$ or $\epsilon^\alpha(t,\sigma)$  into a slot that requires a covector. For instance, considering a spacetime geometry given by  a $(1,2)$-tensor field $G$, one obtains eight tensors of various valences on the manifold $\Sigma$, one of which is the $(0,1)$-tensor field
\begin{equation}\label{dofprojection}
  \mathbf{g}^{0}{}_{0 \alpha}(t,\sigma) := G_{X(t,\sigma)}(\epsilon^{0}(t,\sigma),e_{0}(t,\sigma),e_{\alpha}(t,\sigma))\,,
\end{equation}
which generically differs from the correspondingly defined $\mathbf{g}^{0}{}_{\alpha 0}(t,\sigma)$, which is why we do not suppress the $0$-indices in the notation. It is economical to define one single hyperindex that collects all index combinations for all resulting tensors on $\Sigma$, in some chosen order, such as
\begin{equation}
\mathscr{A}=({}^{0}{}_{00},{}^{0}{}_{0\beta_2},{}^{0}{}_{\beta_10},{}^{0}{}_{\beta_1\beta_2}, {}^{\alpha}{}_{00},{}^{\alpha}{}_{0\beta_2},{}^{\alpha}{}_{\beta_10},{}^{\alpha}{}_{\beta_1\beta_2})\,
\end{equation}
for our example. Note that we abstain from employing potential algebraic symmetries of the spacetime geometry $G$, such as $G^{a}{}_{bc} = G^{a}{}_{[bc]}$, which of course could be used to remove redundant information from the list $\mathbf{g}^\mathscr{A}$. These are most efficiently dealt with later, when identifying the canonical degrees of freedom of the geometry on the manifold $\Sigma$. 

Thirdly, we project the principal tensor field $P$ from spacetime $M$ to the manifold $\Sigma$, resulting in the $\deg P+1$ tensor fields 
\begin{equation}
  \mathbf{p}^{\alpha_1 \dots \alpha_i}(t,\sigma) := {P}_{X(t,\sigma)}(\epsilon^{\alpha_1}(t,\sigma),\dots,\epsilon^{\alpha_i}(t,\sigma), \epsilon^0(t,\sigma), \dots, \epsilon^0(t,\sigma))
\end{equation}
for $i=0,\dots,\deg P$, 
where the total symmetry of $P$ enables the simpler index notation chosen here for the various induced tensor fields $\mathbf{p}$. Due to the definition of the dual projection frame $\epsilon^0, \dots, \epsilon^3$, however, the first two fields of this set are trivial, 
\begin{equation}\label{pframeconditions}
\mathbf{p}(t,\sigma) = 1\qquad\text{and}\qquad \mathbf{p}^\alpha(t,\sigma) = 0\,.
\end{equation}
Finally note that for any fixed value of the foliation parameter $t$, all fields $\mathbf{p}(t)$ and $\mathbf{g}(t)$ present not only tensor fields on $\Sigma$, but, at the same time, are functionals of the embedding map $X_t$. This will become technically relevant in the following subsection.
 
\subsection{\new{General hypersurface deformation algebra}}\label{subsec_hypdef}
The kinematical information, encoded in the triple $(M,G,P)$ in general and the therefrom derived Legendre maps $\ell_x$ in particular, takes its most useful form in the so-called deformation algebra of hypersurfaces.

In order to obtain the latter, consider the functional differential operators 
\begin{equation}
\mathbf{H}_t(n) := \int_\Sigma \mathrm d^3z \, n(z) e^a_0(t,z) \frac{\delta}{\delta X_t^a(z)} \quad\textrm{ and } \quad \mathbf{D}_t(\vec{n}) := \int_\Sigma \mathrm d^3z\, n^\alpha(z) e^a_\alpha(t,z) \frac{\delta}{\delta X_t^a(z)}\,,
\end{equation}
for arbitrary test functions $n$ and $\vec{n}$ on the manifold $\Sigma$, which act on functionals of the embedding maps $X_t: \Sigma \hookrightarrow M$ introduced in the previous subsection. Their geometric meaning, namely as normal and tangential deformation operators, is revealed by letting $n:=\mathbf{n}$ and $\vec{n}:=\vec{\mathbf{n}}$ for the lapse and shift fields $\mathbf{n}$ and $\vec{\mathbf{n}}$ induced by the foliation, see \cite{WittePhD}. The only kinematical information entering here is the Legendre maps $\ell_x$, namely implicitly through definition (\ref{ezero}) of the normal vector field $e_0$ along the hypersurface $X_t(\Sigma)$. 
It is useful to note that the operators $\mathbf{H}_t(n)$ and $\mathbf{D}_t(\vec{n})$ are vector fields on the infinite-dimensional manifold of embeddings $\textrm{Emb}(\Sigma, M)$, for which one can therefore calculate the Lie brackets between vector fields,
\begin{eqnarray}
\big[\mathbf{H}_t(n), \mathbf{H}_t(m)\big] &=& - \mathbf{D}_t((\text{deg\,}P-1) \mathbf{p}_t^{\alpha\beta}(m\partial_{\beta}n - n\partial_{\beta}m)\partial_{\alpha}) \,, \label{eq:deform1}\\ 
\big[\mathbf{D}_t(\vec{n}), \mathbf{H}_t(m)\big] &=& - \mathbf{H}_t(\mathcal{L}_{\vec{n}}m) \,, \label{eq:deform2}\\
\big[\mathbf{D}_t(\vec{n}), \mathbf{D}_t(\vec{m})\big]  &=& - \mathbf{D}_t(\mathcal{L}_{\vec{n}}\vec{m}) \label{eq:deform3}\,.
\end{eqnarray} 
Note that the right hand side of (\ref{eq:deform1}) depends on the component functions $\mathbf{p}^{\alpha\beta}$ of the induced principal polynomial (but not any $\mathbf{p}_t^{\alpha_1\dots\alpha_n}$ with $n\neq2$) and thus on the initially specified matter field dynamics and their geometric background; this is indeed the only, but all-important trace left in the algebra by the Legendre maps.

The failure of these Lie brackets to close with only structure constants, rather than structure functions, has a number of complicating implications. 
Chief among those is that one cannot simply represent the above relations --- at least not without a number of additional requirements such as those we will make in section \ref{subsec_HKT} --- as a Lie algebra of functionals of some geometric phase space variables.

\subsection{Canonical geometry}\label{subsec_primary}
\new{We now revert the perspective taken in the two preceding subsections, where the spacetime geometry was considered as primarily given and the induced geometry on the leaves of some foliation as a derived, thus  secondary, quantity. Indeed, the canonical point of view taken here now considers the geometry on the leaves as primary and the spacetime geometry as only reconstructed from there by virtue of the foliation. 
This change of perspective comes at the price that four generically non-linear conditions, which the induced geometry 
satisfied by construction, must now be reinstated explicitly for the canonical geometry. 
}

More precisely, the transition from the induced geometry to the canonical one proceeds as follows. If the geometry  $\mathbf{g}^\mathscr{A}(t)$ is induced from a spacetime geometry $G$ by virtue of a foliation $X_t: \Sigma \hookrightarrow M$, together with an induced lapse $\mathbf{n}(t)$ and induced shift $\mathbf{n}^\alpha(t)$, then we introduce 
\begin{equation}
g^\mathscr{A}(t)\,, \qquad n(t)\,,\qquad n^\alpha(t)
\end{equation}
as new, independent one-parameter families of tensor fields on $\Sigma$, which capture precisely the tensor structure of the fields $\mathbf{g}^\mathscr{A}(t)$, the lapse $\mathbf{n}(t)$ and the shift $\mathbf{n}^\alpha(t)$. Note  that the construction of the induced tensor fields $\mathbf{g}^\mathscr{A}(t)$ automatically equips them with properties that are not captured by their mere tensor valence, while their valence is indeed the only information left after the transition to $g^\mathscr{A}(t)$.
How to reinstate the missing information will be remedied in the next subsection. This will lead directly to the associated bundle techniques mentioned above. We also need to translate any quantities that were previously defined in terms of the induced geometry $\mathbf{g}^\mathscr{A}$, into corresponding quantities of the $g^\mathscr{A}$. The most relevant such transition for the purposes of this paper, is the one from the $\mathbf{p}^{\alpha_1\dots\alpha_i}(t)$ to the new one-parameter families of fields 
$p^{\alpha_1\dots\alpha_i}(t)$ for $i=0,\dots,\deg P$,
which are defined as precisely the same functions of $g^\mathscr{A}(t)$ as the $\mathbf{p}^{\alpha_1\dots\alpha_i}(t)$ were of the $\mathbf{g}^\mathscr{A}(t)$. 

The most relevant property of the induced geometry $\mathbf{g}^\mathscr{A}$, which is not automatically captured by the canonical geometry  $g^\mathscr{A}$, is the frame conditions (\ref{pframeconditions}). While these are satisfied
for the induced fields $\mathbf{p}$ and $\mathbf{p}^\alpha$ by construction, this information is lost when the functionals $\mathbf{g}$ are replaced by the fields $g^\mathscr{A}$ that merely mimic their tensorial structure. Thus the normalization and annihilation conditions must be explicitly reinstated as
\begin{equation}\label{projectionconditions}
p(g)(t) = 1 \qquad \textrm{ and }\qquad p^\alpha(g)(t)=0\,.
\end{equation} 
These conditions impose four --- generically non-linear --- conditions on the canonical geometry $g^\mathscr{A}$ and thus effectively remove four of their  degrees of freedom. These non-linear relations are captured, beginning with the next subsection, by a suitable parametrization of the $g^\mathscr{A}$.

Similarly, any algebraic symmetry of the spacetime geometry $G$ is automatically passed on to the induced tensor fields $\mathbf{g}^\mathscr{A}$, but must again be explicitly reinstated for the canonical geometry $g^\mathscr{A}$ by additional, now however linear and homogeneous, conditions 
\begin{equation}\label{symmetryconditions}
 (\delta^\mathscr{A}_\mathscr{B} - \Pi^\mathscr{A}{}_\mathscr{B}) g^\mathscr{B} = 0 \,
\end{equation} 
for suitable projectors $\Pi$. These additional symmetry conditions can be implemented without extra effort alongside the generically non-linear frame conditions by the method developed in the following subsection. \new{The possibility of such a combined treatment was the conceptual reason for withholding the implementation of symmetry conditions before. An independent technical reason for not reflecting algebraic symmetries already at the level of the $g^\mathscr{A}$ was to be able to calculate partial derivatives of functions depending on the $g^\mathscr{A}$, which requires to be able to vary each individual entry while keeping all others fixed.}

\subsection{Parametrization of the canonical geometry}\label{subsec_generalized}	
The configuration variables of the gravitational dynamics, which we are about to construct, parameterize, without further constraints, canonical geometries $g^\mathscr{A}$ that respect the frame and symmetry conditions identified in the previous subsection. 
But because of their generic overall non-linearity, these conditions cannot be implemented by simply cutting away some tensor field components among the $g^\mathscr{A}$ while keeping others. In fact, the situation is pretty much the same as for a particle in Euclidean space that is conditioned to move on an embedded submanifold, such as a circle. One cannot simply cut away one of the Cartesian coordinates, as one could if the particle was constrained to a linear subspace instead. The conceptually and technically best solution in classical mechanics is to introduce generalized coordinates. The same idea applies here.  
We require exactly as many field configuration variables $\varphi^1,\dots, \varphi^F$ as are needed to bijectively parametrize the tensor fields  $g^\mathscr{A}$ such that the frame conditions (\ref{projectionconditions}) and symmetry conditions (\ref{symmetryconditions}) are met by construction.   
Technically, this is achieved by choosing a suitable $F$-dimensional manifold $\Phi$ and smooth maps 
$\widehat g^\mathscr{A}: \Phi \to \mathbb{R}$
such that any canonical geometry $g^\mathscr{A}$ generated by $\widehat g^\mathscr{A}(\varphi^1,\dots,\varphi^F)$ satisfies the conditions
\begin{equation}
 (\delta^\mathscr{A}_\mathscr{B} - \Pi^\mathscr{A}{}_\mathscr{B}) {\widehat g}^\mathscr{B}(\varphi(t)) = 0\,\quad\textrm{ and }\quad
p(\widehat g(\varphi(t,\sigma))) = 1 \quad\textrm{ and } \quad p^\alpha(\widehat g(\varphi(t,\sigma))) = 0 
\end{equation}
for any $\sigma\in\Sigma$ and any real $t$ in the range of the foliation parameter. If one single map $\widehat g^\mathscr{A}$ does not suffice to cover the required range, the usual chart transition constructions can be invoked. The number $F$ of configuration variables is the total number of all  $g^\mathscr{A}$ minus the normalization condition minus the three annihilation conditions and minus the dimension of the eigenspace $\textrm{Eig}_1(\Pi)$ of the projector $\Pi$. For instance, when the triple $(M,G_\text{\tiny metric},G_\text{\tiny metric}^{-1})$ induced by a Lorentzian metric $G_\text{\tiny metric}$, there are $F= 16-3-1-6 = 6$ configuration field variables, which, due to all constraints being linear in this case, can coincidentally be written as a not further constrained metric tensor on the three-dimensional manifold $\Sigma$.

Conversely, we require the existence of inverse maps $\widehat \varphi^A$ that send any collection $g^\mathscr{A}$ (even if the frame and symmetry conditions are not met) to a real number, but which are constructed such that
\begin{equation}
 (\widehat \varphi^A\circ\widehat g)(\varphi) = \varphi^A\qquad \textrm{ for } A = 1, \dots, F\,.
\end{equation}
The opposite composition $(\widehat{g}^\mathscr{A}\circ\widehat{\varphi})$ projects any set of $g^\mathscr{A}$, even if the latter does not yet satisfy the frame and symmetry conditions, to a set that does. Ubiquitous appearance throughout the theory is then made by the maps
\begin{equation} 
\frac{\partial\widehat{\varphi}^A}{\partial g^{\mathscr{A}}}(\widehat g(\varphi)) \qquad \textrm{and}\qquad
	\frac{\partial\widehat{g}^\mathscr{A}}{\partial \varphi^A}(\varphi)\,,
\end{equation}
as they emerge as intertwiners between the components of the canonical geometry, labeled by $\mathscr{A}$, and the components of the configuration variables, labeled by $A$. The above defining conditions for the maps $\widehat g$ and $\widehat \varphi$ immediately imply the important and heavily used completeness relations
\begin{equation}
\frac{\partial\widehat{\varphi}^A}{\partial g^{\mathscr{A}}}(\hat g(\varphi)) \frac{\partial\widehat{g}^\mathscr{A}}{\partial \varphi^B}(\varphi) = \delta^{A}_{B}  
\qquad \text{and} \qquad 
\frac{\partial\widehat{g}^\mathscr{A}}{\partial \varphi^A}(\varphi) \frac{\partial\widehat{\varphi}^A}{\partial g^{\mathscr{B}}}(\hat g(\varphi)) = \mathscr{T}^{\mathscr{A}}{}_{\mathscr{B}}(\varphi) \,,
\end{equation}
where $\mathscr{T}^{\mathscr{A}}{}_{\mathscr{B}}(\varphi)$ is defined by the left hand side and is easily seen to be a projector.

\section{Canonical gravitational dynamics}\label{sec_canonicaldynamics}
Employing the technology developed in the previous section, we now significantly improve and extend the results of \cite{GSWW}. The crucial technical advance is the identification of the geometric phase space with the non-tensorial configuration variables and canonically conjugate 
momentum densities, whose transformation behavior already captures the non-linear constraints on the canonical geometry that was left to be implemented only afterwards in previous treatments. The complete determination of the gravitational Hamiltonian, 
which is required to satisfy the two properties imposed in section \ref{subsec_HKT}, 
then finally leads to the gravitational closure equations for any matter field dynamics that satisfy the matter conditions from subsection \ref{subsec_pdr}. This is a countably infinite set of partial differential equations that needs to be solved in order to obtain the gravitational Hamiltonian or, equivalently, Lagrangian density.     

\subsection{Canonical phase space}\label{subsec_phasespace}
\label{section_phase_space}
Having identified the unconstrained geometric configuration variables $\varphi^A$ for a spacetime geometry $(M,G,P)$ in the previous section,
we are now in the position to adjoin canonically conjugate momentum fields $\pi_A$ with respect to the field-theoretic Poisson bracket
\begin{equation}\label{Poissonbracket}
\left\{F,G\right\} := \int_\Sigma \mathrm d^3z \left(\frac{\delta F}{\delta \varphi^A(z)} \frac{\delta G}{\delta \pi_A(z)}  -   \frac{\delta G}{\delta \varphi^A(z)} \frac{\delta F}{\delta \pi_A(z)} \right)\,,
\end{equation}
which is to be evaluated on any two scalar functionals $F[\varphi,\pi]$ and $G[\varphi,\pi]$ of the canonical configuration variables $\varphi^A$ and the associated canonical momenta $\pi_A$. \new{We remark in passing that, as usual, there is an ambiguity in the choice of the canonical momenta for some given set of configuration variables $\varphi^A$. For if $\pi_A$ presents a possible choice, then so does $\pi_A + \Lambda_A[\varphi]$ for any closed covector field $\Lambda_A \delta \varphi^A$ on configuration space that satisfies the closure condition}
\begin{equation}
  \frac{\delta\Lambda_A}{\delta \varphi^B} -  \frac{\delta\Lambda_B}{\delta \varphi^A} = 0\,.
\end{equation}
 
From the obvious requirement that the bracket (\ref{Poissonbracket}) be well-defined under changes of coordinate chart on the manifold $\Sigma$, we can derive the precise mathematical nature of the momenta. Technically, the key observation
is that the $F$ configuration variables $\varphi^A$ are a section of an $F$-dimensional $\Phi$-fibre bundle over $\Sigma$, which is an associated bundle with respect to the frame bundle $L\Sigma$ by virtue of the (generically non-linear) group action $\rho: GL(3) \times \Phi \to \Phi$ that is enforced by the way the $\varphi^A$  transform under coordinate transformations, namely
\begin{equation}
\rho^A\!\left(\tfrac{\partial \widetilde z}{\partial  z}, \varphi\right) := \widehat{\varphi}^A \left( \mathcal{R}^{\mathscr{A}}{}_{\mathscr{B}}\!\left(\tfrac{\partial \widetilde z}{\partial  z}\right)\, \widehat{g}^{\mathscr{B}}(\varphi^1,\dots\varphi^F)  \right)\,, 
\end{equation}
where $\mathcal{R}^{\mathscr{A}}{}_{\mathscr{B}}\!\left(\tfrac{\partial \widetilde z}{\partial  z}\right)$ denotes the standard tensorial action of the $GL(3)$-transformation $\partial \widetilde z/\partial  z$ on the various tensors on $\Sigma$ which we collectively labeled by $\mathscr{B}$.
Note that the above transformation behavior of configuration variables is not a postulate, but directly follows from our choice of parameterization map $\widehat \varphi$ and its inverse $\widehat g$, on which the group action then naturally depends. But with the transformation behavior of the configuration variables thus under control, we can now straightforwardly read off the $GL(3)$ group action that defines a further associated $\Pi$-fibre bundle over the manifold $\Sigma$, of which the canonical momenta $\pi_A$ shall constitute a section. In order for the Poisson bracket above to be well-defined, we then need to impose the group action $\rho^*: GL(3) \times \Pi \to \Pi$
\begin{equation}
\rho_A^*\left(\tfrac{\partial \widetilde z}{\partial  z},\pi\right) :=  \left(\det \frac{\partial \widetilde z}{\partial z}\,\right)\,  \frac{\partial \widehat \varphi^B}{\partial g^{\mathscr{A}}} (\mathcal{R}^{-1}){}^{\mathscr{A}}{}_{\mathscr{B}}\!\left(\tfrac{\partial \widetilde z}{\partial  z}\right)  \frac{\partial \widehat g^{\,\mathscr{B}}}{\partial \varphi^A} \widetilde \pi_B\,.
\end{equation}
Indeed, it is easy to see that then the Poisson bracket is well-defined, because the functional derivative $\delta F/\delta \varphi^A(z)$ has density weight one (since $\varphi^A$ has density weight zero), while the fact that $\pi_A$ already has density weight one cancels the density weight from the functional differentiation in $\delta G/\delta \pi_A(z)$, rendering the latter of weight zero. Thus the integrand of the Poisson bracket can be shown to be a scalar density of weight one and thus the integral to be well-defined.

\subsection{\new{Embedding properties and gravitational Hamiltonian}}\label{subsec_HKT}
In this section, we introduce two functionals on the just constructed phase space, whose action on the configuration variables mimics the action of the normal and tangential deformation operators of section \ref{subsec_hypdef} on the geometry projected to the leaves of a given spacetime foliation, and then formulate two {\it embedding properties} concerning the spacetime interpretation of these  canonical objects. 

\begin{center}{\it Embedding property 1: Local phase space avatars of deformation operators}\end{center}
We require that there are phase space functionals
\begin{equation}\label{curlHcurlD}
\mathscr{H}(n) := \int_\Sigma \mathrm d^3z \, n(z) \mathcal{H}[\varphi(z),\pi(z)]  \quad\textrm{ and }\quad \mathscr{D}(\vec{n}) := \int_\Sigma \mathrm d^3z \,n^\alpha(z) \mathcal{D}_\alpha[\varphi(z),\pi(z)]
\end{equation}
in terms of local functionals $\mathcal{H}$ and $\mathcal{D}$ of the geometric phase space variables, 
which evolve the canonical data between leaves of a given spacetime foliation $X_t$ such that the result agrees with what the application of the normal and tangential deformation operators $\mathbf{H}_t(n)$ and $\mathbf{D}_t(\vec{n})$ yield when they are applied to the projected geometry,
\begin{equation}\label{Michel}
\mathbf{H}_t(n) \mathbf{g}_t^\mathscr{A} = - \big\{\mathscr{H}(n), g^\mathscr{A}\big\}_t \qquad\textrm{and}\qquad\mathbf{D}_t(\vec{n}) \mathbf{g}_t^\mathscr{A} = -\big\{\mathscr{D}(\vec{n}), g^\mathscr{A}\big\}_t\,,
\end{equation}
where the equal signs are to be understood in the sense that the right hand side is the same function of the canonical geometry $g^\mathscr{A}$ as the left hand side is of the induced geometry $\mathbf{g}^\mathscr{A}$. Now in order to ensure the consistency of the spacetime interpretation of the action of $\mathscr{H}(n)$ and $\mathscr{D}(\vec{n})$, as it is expressed in terms of the two requirements (\ref{Michel}), we additionally stipulate the Poisson algebra 
\begin{eqnarray}
\{ \mathscr{H}(n), \mathscr{H}(m)  \} &=& \mathscr{D}\left((\text{deg }P-1) p^{\alpha\beta}(m\partial_{\beta}n - n\partial_{\beta}m)\partial_{\alpha}\right) \,, \label{eq:poisson1}\\
\{ \mathscr{D}(\vec{n}) , \mathscr{H}(m)  \} &=& \mathscr{H}(\mathcal{L}_{\vec{n}}\,m) \,, \label{eq:poisson2}\\
\{ \mathscr{D}(\vec{n}) , \mathscr{D}(\vec{m})  \} &=& \mathscr{D}(\mathcal{L}_{\vec{n}}\,\vec{m}) \label{eq:poisson3}\,,
\end{eqnarray}
which ensures that there is no inconsistency with the vector field algebra (\ref{eq:deform1})--(\ref{eq:deform3}) of deformation operators $\mathbf{H}(n)$ and $\mathbf{D}(\vec{n})$. The five equations (\ref{Michel}) and (\ref{eq:poisson1})--(\ref{eq:poisson3}) will play a crucial r\^ole in all that follows from now on, starting from the below determination of the general form of the Hamiltonian of a spacetime diffeomorphism invariant theory up to the calculation of the gravitational closure equations as the central result of this article. 

\begin{center}{\it Embedding property 2: Spacetime diffeomorphism invariance and path-independence}\end{center}
Spacetime diffeomorphism invariance of the canonical dynamics certainly requires that the evolution of initial data between any two fixed Cauchy surfaces be independent of the choice of intermediate foliation. Such path-independence of the dynamics implies (see section 5 of \cite{HKT} for the general line of argument that applies mutatis mutandis in our general setting) that the pertinent Hamiltonian density must be of the totally constrained form 
\begin{equation}\label{HamHD}
H[\varphi,\pi;n,\vec{n}) =  \mathscr{H}(n) + \mathscr{D}(\vec{n})\,,
\end{equation}
whence the functionals (\ref{curlHcurlD}) are commonly referred to as the superhamiltonian and supermomentum {\it constraints}. The closure of the constraint algebra (\ref{eq:poisson1})--(\ref{eq:poisson3}) ensures that the Hamiltonian density (\ref{HamHD}) does not give rise to further constraints and thus does not pick up additional terms.

We briefly remark on a well-known subtlety with regards to the intimate relationship between path-independence and spacetime diffeomorphism invariance of canonical dynamics. 
Indeed, while the required path-independence of the canonical dynamics geometrically implements the idea that the dynamics be invariant under spacetime diffeomorphisms, at first sight it may be somewhat disturbing to learn that not even infinitesimal spacetime diffeomorphisms can be represented on our geometric phase space constituted by the $\varphi^A$ and $\pi_A$. This can be remedied, however, by first extending the phase space such as to additionally include the four  component fields $X^a$ of the foliation map and associated canonically conjugate momenta $\Pi_a$ and then dressing up the Hamiltonian (\ref{HamHD}) such as to include these new variables. Exactly along the same lines demonstrated in \cite{IshamKuchar1} for parametrized dynamics and in \cite{IshamKuchar2} for the case of a metric geometric phase space, one then constructs an action of the diffeomorphism algebra on the extended phase space. Since these steps do not change the physical contents of the theory, 
the issue of understanding the diffeomorphism invariance of the gravitational theories obtained by gravitational closure is resolved in same fashion as in standard general relativity. 

\subsection{Functional differential reformulation of the constraint algebra}\label{subsec_fdform}
The conditions (\ref{Michel}) and the constraint algebra (\ref{eq:poisson1})--(\ref{eq:poisson3}) provide functional differential conditions on $\mathscr{H}(n)$ and $ \mathscr{D}(\vec{n})$  that turn out to be so strong as to determine the gravitational superhamiltonian and supermomentum \new{under the matter conditions listed in \ref{subsec_pdr} and the embedding properties of the Hamiltonian stipulated in \ref{subsec_HKT}}. The resulting Hamiltonian (\ref{HamHD}) then generates the evolution of phase space curves $(\varphi^A(t), \pi_A(t))$ with respect to what, from a spacetime point of view, is the foliation parameter $t$. The thus generated ``geometry movie'' on the manifold $\Sigma$ can then be embedded, frame by frame, into the spacetime manifold by virtue of the one-parameter family $X_t: \Sigma \hookrightarrow M$ by letting $n:=\mathbf{n}$ and $\vec{n}:=\vec{\mathbf{n}}$, which results in the immutable spacetime geometry $G$. This is the mechanism underlying the dynamical closure of prescribed canonically quantizable matter dynamics. In this subsection, we solve the autonomous third constraint equation (\ref{eq:poisson3})  for the supermomentum $\mathscr{D}(\vec{n})$ and are thus able to reformulate the first two constraint equations (\ref{eq:poisson1}) and (\ref{eq:poisson2}) as linear functional differential equations for a suitable Lagrangian functional. 

Carefully taking into account the parametrization $\widehat g^\mathscr{A}$ of the canonical geometry $g^\mathscr{A}$ in terms of the configuration variables $\varphi^A$, one finds that the second condition of (\ref{Michel}) together with the constraint algebra equation (\ref{eq:poisson3}) already completely determines the constraint functional to be
\begin{equation}
\mathscr{D}(\vec{n}) = \int_{\Sigma} \mathrm{d}^3z  \, \pi_A(z) \,\frac{\partial\widehat\varphi^A}{\partial g^\mathscr{A}}(\widehat g(\varphi(z))) \,\left(\mathcal{L}_{\vec{n}} \, \hat{g}(\varphi)\right)^{\mathscr{A}}(z) \,,
\end{equation}
with the only, but significant, improvement compared to (4.16) of \cite{GSWW} being  the appearance of the intertwiner map $\partial\widehat\varphi^A/\partial g^\mathscr{A}$ and the parametrization map $\widehat g^\mathscr{A}$.

\new{The first condition of (\ref{Michel}), in contrast, is much weaker, for it does not determine in any way the superhamiltonian's ultralocal dependence on the momentum fields $\pi_A$. But it crucially allows to determine the dependence on {\it derivatives} of the momentum fields. Indeed, as is shown in \cite{WittePhD}, it enforces
\begin{equation}\label{eq:partialHam}
\mathscr{H}(n) = \int_\Sigma \mathrm d^3z\, n(z)\left[\mathcal{H}_\textrm{\tiny local}[\varphi;\pi)(z) - \partial_\gamma\left( M^{A\,\gamma}(\varphi) \pi_A\right)(z)   \right]\,,
\end{equation}
for a still entirely unknown functional $\mathcal{H}_{\tiny local}[\varphi,\pi)$ that is local in the configuration variables and ultralocal in the momenta. The dependence on derivatives of the momenta in equation (\ref{eq:partialHam}) is controlled by the coefficient
\begin{equation}
M^{A\,\gamma}(\varphi) := \frac{\partial\widehat\varphi^A}{\partial g^{\mathscr{A}}}(\widehat g(\varphi)) \, e_0^a(t,\sigma) \frac{\partial\mathbf{g}^\mathscr{A}}{\partial\partial_\gamma X^a}(t,\sigma)\,,
\end{equation}
whose last factor is easily calculated in practice} from the definition of the $\mathbf{g}^\mathscr{A}$ using the identities
\begin{equation}
  \frac{\partial e_0^m}{\partial\partial_\gamma X^a} = -(\deg P - 1)e^m_\sigma e_a^0 \mathbf{p}^{\sigma\gamma} \qquad \textrm{ and } \qquad
   \frac{\partial e^m_\mu}{\partial\partial_\gamma X^a} = \delta^m_a \delta^\gamma_\mu 
\end{equation}
for the tangent frame fields, and
\begin{equation}
 \frac{\partial \epsilon^0_m}{\partial\partial_\gamma X^a} = - \epsilon^0_a \epsilon^\gamma_m \qquad \textrm{ and } \qquad
  \frac{\partial \epsilon^\mu_m}{\partial\partial_\gamma X^a} = - \epsilon^\mu_a \epsilon^\gamma_m  + (\deg P -1) \epsilon^0_m \epsilon^0_a \mathbf{p}^{\mu\gamma}\,
\end{equation}
for the cotangent fields. 

Thus the gravitational Hamiltonian (\ref{HamHD}) is determined so far only up to the functional $\mathcal{H}_\textrm{\tiny local}[\varphi;\pi)$. The determination of this remaining piece of the superhamiltonian requires significant work and will finally lead to the gravitational closure equations in the next subsection. 

We prepare the derivation of these closure equations by following again \cite{GSWW} closely in applying a trick due to Kucha\u r \cite{Kuchar1974}, which converts both the first two constraint equations into linear equations.
To this end, we define the generalized velocity fields
\begin{equation}
  k^A[\varphi;\pi) := \frac{\partial\mathcal{H}_\textrm{\tiny local}}{\partial \pi_A}[\varphi;\pi)
\end{equation}
and subsequently perform a formal Legendre transformation on the $\pi_A$, rewriting  
\begin{equation}
\mathcal{H}_\textrm{\tiny local}[\varphi;\pi) = \pi_A k^A[\varphi;\pi) - \mathcal{L}[\varphi;k[\varphi;\pi)) 
\end{equation}
and thus trading the unknown density $\mathcal{H}_\textrm{\tiny local}[\varphi,\pi)$ for another unknown density $\mathcal{L}[\varphi,k)$. The benefit of this trade, however, is that the quadratic condition (\ref{eq:poisson1}) on $\mathscr{H}$ is converted into a merely linear homogeneous functional differential equation for the functional $\mathcal{L}$. Indeed, using the same idea as in \cite{GSWW}, but now employing the parameterization $\widehat g^\mathscr{A}$ of the canonical geometry in terms of the configuration variables $\varphi$, one picks up crucial additional terms and finally obtains the functional differential form of (\ref{eq:poisson1}). More precisely, using the 
\begin{equation}
       \textrm{shorthand notation }\quad  Q_{:A}{}^{\alpha_1\dots\alpha_N} := \frac{\partial Q}{\partial \partial_{\alpha_1\dots\alpha_N}\varphi^A}\,,
\end{equation}
to denote partial derivatives with respect to partial derivatives of configuration variables over $\Sigma$, where $Q$ is any differentiable function of the configuration variables and their partial derivatives on $\Sigma$, this functional differential equation equivalent to the bracket (\ref{eq:poisson1}) reads 
\begin{eqnarray}
	0 &=& - k^B(y) \left( \frac{\delta \mathcal{L}(x)}{\delta\varphi^B(\cdot)} \right)(y) + \left(\partial_\gamma \delta_x\right)(y) k^B(y) {M^{A\gamma}{}_{:B}}(x) \frac{\partial \mathcal{L}}{\partial k^A}(x) + \partial_\mu \left( \frac{\delta \mathcal{L}(x)}{\delta\varphi^B(\cdot)} {M^{B\mu}{}} \right)(y) \nonumber \\
	& & + \partial_\mu\left( \frac{\partial \mathcal{L}}{\partial k^A} \right)\!(x) \Big[  (\mathrm{deg} P-1){p^{\rho\mu}}{F^A{}_\rho{}^{\nu}} - {M^{B[\mu|} M^{A|\nu]}{}_{:B}} \Big](x) \left(\partial_\nu\delta_x\right)(y) \nonumber\\
	& & - \frac{\partial \mathcal{L}}{\partial k^A}(x) \Big[ (\mathrm{deg} P-1) {p^{\rho\nu}}\left({E^A{}_\rho} + {F^A{}_\rho{}^\gamma{}_{,\gamma}}\right)  + \partial_\mu \left( {M^{B[\mu|}}{M^{A|\nu]}{}_{:B}} \right) \Big] (x) \left(\partial_\nu\delta_x\right)(y) \nonumber\\ 
    & &- (x \longleftrightarrow y)	\label{functionaleqn1}
\end{eqnarray}
with the new coefficients $E^A{}_\mu$ and $F^A{}_\mu{}^{\nu}$ defined by 
\begin{equation}
  \frac{\partial \widehat \varphi^A}{\partial g^\mathscr{A}} \left(\mathcal{L}_{\vec{n}} \widehat g(\varphi)\right)^\mathscr{A} =:  n^\mu E^A{}_\mu - \partial_\gamma n^\mu F^{A}{}_\mu{}^\gamma \,.
\end{equation}
Similarly, one rewrites the second constraint algebra relation \eqref{eq:poisson2} as an additional linear homogeneous functional differential equation for the density $\mathcal{L}$, namely 
\begin{eqnarray}
	0 &=& \left( \frac{\partial \mathcal{L}}{\partial k^B} \right)\!\!(y)\, k^A(y) \Big[ {E^B{}_{\mu:A}}(y) \delta_y(x) + \left({E^B{}_{\mu:A}{}^\gamma} + {F^B{}_{\mu}{}^\gamma{}_{:A}} \right)(y) (\partial_\gamma \delta_y)(x) \Big] \nonumber \\
    & & - k^A(y) ( \partial_\gamma \frac{\partial \mathcal{L}}{\partial k^B})(y) {F^B{}_{\mu}{}^\gamma{}_{:A}} (y) \delta_y(x) \nonumber \\
    & & - \left( k^A  \frac{\partial \mathcal{L}}{\partial k^A} - \mathcal{L}\right)(y) (\partial_\mu \delta_y)(x) + \partial_\mu \left( k^A  \frac{\partial \mathcal{L}}{\partial k^A} - \mathcal{L} \right)(y) \delta_y(x) \nonumber \\
    & & + \left({E^A{}_\mu} + {F^A{}_\mu{}^\gamma{}_{,\gamma}}\right) (x) \frac{\delta \mathcal{L}(y)}{\delta \varphi^A(x)} + {F^A{}_\mu{}^\gamma} (x) \partial_\gamma \left( \frac{\delta \mathcal{L}(y)}{\delta \varphi^A(\cdot)} \right)(x)\,.\label{functionaleqn2}
\end{eqnarray}
The coefficients $E^A{}_\mu$, $F^A{}_\mu{}^{\nu}$, $M^{B\mu}{}$ and $p^{\alpha\beta}$ are completely determined by the triple $(M,G,P)$, and need to be provided as input when solving the functional differential equations (\ref{functionaleqn1}) and (\ref{functionaleqn1}), or the indeed equivalent closure equations derived in the next subsection, for the only remaining unknown functional $\mathcal{L}$. We will therefore refer to these four types of coefficients as the {\it input coefficients} from now on. They are always directly calculated from the initially specified matter dynamics and their background geometry $G$.   

\subsection{Gravitational closure equations}\label{subsec_L}\label{subsec_towardsconstruction}
\new{The gravitational closure equations for any given field theory satisfying the matter conditions of \ref{subsec_pdr} are the countably infinite set of {\it partial differential equations} that follow for the sequence of coefficient functionals
\begin{equation}
C[\varphi]\,,\quad C_{A_1}[\varphi]\,,\quad C_{A_1 A_2}[\varphi]\,,\quad\dots,,
\end{equation}
which depend at most locally on the configuration variables $\varphi^A$,
upon insertion of the functional 
\begin{equation}
\mathcal{L}[\varphi;K) =  \sum_{N=0}^\infty {C_{A_1\dots A_N}}[\varphi] k^{A_1} \cdots k^{A_N}
\end{equation}
into the functional differential equations (\ref{functionaleqn1}) and (\ref{functionaleqn2}). This reformulation of the two functional differential equations comes at the price of now having to solve countably many equations which however makes the problem directly accessible to the full machinery  \cite{SeilerBook} that is nowadays available for the study of systems of linear partial differential equations.}

The derivation of these linear homogeneous equations, which present the desired gravitational closure equations, is a painstaking exercise. Despite two crucial modifications, it proceeds technically in full analogy to the steps performed in \cite{GSWW}. The first modification presented by our now employing a parametrization $\widehat g^\mathscr{A}$ of the canonical geometry $g^\mathscr{A}$ in terms of the unconstrained configuration variables $\varphi^A$, such that the generically non-linear polynomial frame conditions and any additional symmetry conditions for the tensor fields $g^\mathscr{A}$ are captured automatically. The second modification is that we now convert also the second functional differential equation for $\mathcal{L}$, equation (\ref{functionaleqn2}), into a set of partial differential equations, since the (anyhow somewhat awkward) workaround taken before is no longer available for the generalized tensor components we now use as configuration degrees of freedom \cite{NadineMSc}. Since it is ultimately straightforward to adapt the calculations of \cite{GSWW} to the new technical developments of this paper, we content ourselves with displaying the resulting set of linear homogeneous partial differential equations in terms of the
$$\textrm{seven individual equations } (C1) \textrm{ to } (C7) \textrm{ and}$$
$$\textrm{fourteen sequences of equations } (C8_{N}) \textrm{ to } (C21_{N}) \textrm{ for } N\geq 2$$ 
on the last two landscape pages of this article. Those are the {\it gravitational closure equations}. Set up by provision of the matter-determined {\it input coefficients} $E^A{}_\mu$, $F^A{}_\mu{}^{\nu}$, $M^{A\mu}{}$ and $p^{\alpha\beta}$, their solution yields the sequence of {\it output coefficients} $\left\{C_{A_1\dots A_N}[\varphi]\right\}_{N\geq 0}$ and thus the dynamics for the spacetime geometry.   

The explicit form of these gravitational closure equations, as they are listed on the last two pages of this article, has already been simplified in so far as their derivation yields that, for $N\geq 2$, all output coefficients are functions
\begin{equation}
  C_{A_1 A_2 \dots A_N}(\varphi, \partial\varphi,\partial\partial\varphi) \,,
\end{equation}
which only depend on at most second partial derivatives of the configuration variables $\varphi^A$ with respect to the base manifold $\Sigma$ . A weaker result applies to the first two output coefficients
\begin{equation}
  C[\varphi] \qquad\textrm{ and }\qquad C_A[\varphi] \,,
\end{equation}
namely that if $C_A$ depends on partial derivatives of the $\varphi$ up to the $D$-th order, then $C$ depends on 
partial derivatives up to order $\max\{2,D+1\}$. 
A stronger result holds if the input coefficient $M^{A\mu}$ vanishes identically, for then $C$ depends on the configuration variables $\varphi$ to at most second derivative order.  
Thus one of the first questions one typically wishes to address early on, when solving the gravitational closure equations for specific input coefficients, is the value of $D$.


\subsection{Canonical equations and equivalent spacetime action}\label{sec_spacetimedynamics}
A practically most convenient result is turned up by translation of our results from the canonical picture back to a spacetime formulation. Indeed, the gravitational closure equations immediately provide a perfectly simple, ready-to-use spacetime action that just needs to be varied, as usual, with respect to the components of the spacetime geometry in order to obtain the gravitational field equations.

In the canonical picture, it is the Hamiltonian (\ref{HamHD}) that determines the evolution of our canonical configuration and momentum degrees of freedom according to 
\begin{equation}\label{Hamil}
\dot{\varphi}_t^A(y) = \{ \varphi^A(y), H(n,\vec{n}) \}_t \qquad \text{and} \qquad \dot{\pi}_A(y) = \{ \pi_{A}(y) , H(n,\vec{n}) \}_t \,,
\end{equation} 
where the dot denotes the derivative with respect to the foliation parameter $t$.  The parameter $t$ as well as the lapse $n_t$ and shift $\vec{n}_t$ precisely parametrize the possible choices one could make to embed the three-dimensional manifold $\Sigma$, on which the canonical dynamics play out, into the four-dimensional spacetime. The required diffeomorphism invariance of the theory is precisely the freedom to choose this embedding without changing the contents of the theory, see the first and second embedding property in section \ref{subsec_HKT}.

Inclusion of matter, with a Hamiltonian $H_\text{\tiny matter}[A;\varphi,n,n^\alpha)$ that does not depend on derivatives of the $\varphi^A$, $n$ and $n^\alpha$, thus leads to the geometric evolution equations 
\begin{equation}
\frac{\delta H_\text{\tiny matter}}{\delta \varphi(x)} = - \left[\partial_t - n^\mu \partial_\mu - \partial_\mu n^\mu + (\partial_\gamma n) \frac{\partial M^{B\,\gamma}}{\partial \varphi} - (\partial_\gamma n^\mu)\frac{\partial F{}^B_\mu{}^\gamma}{\partial\varphi^A} \right]\frac{\partial \mathcal{L}}{\partial k^B}(x) + \int \mathrm d^3y \,n(y) \frac{\delta \mathcal{L}(y)}{\delta\varphi(x)}\,, 
\end{equation}
and the two constraint equations 
\begin{equation}
\frac{\delta H_\text{\tiny matter}}{\delta n(x)} = - \left[k^A - \partial_\gamma M^{A\gamma}-M^{A\gamma}\partial_\gamma\right] \frac{\partial\mathcal{L}}{\partial k^A}(x) + \mathcal{L}(x)
\end{equation}
and
\begin{equation}
\frac{\delta H_\text{\tiny matter}}{\delta n^\mu(x)} = - \left[\partial_\mu\varphi^A + \partial_\gamma F{}^A{}_\mu{}^\gamma + F{}^A{}_\mu{}^\gamma \partial_\gamma\right] \frac{\partial \mathcal{L}}{\partial k^A}(x)\,,
\end{equation}
in all three of which the $k^A$ are to be replaced by
\begin{equation}
k^A(x) = \frac{1}{n(x)}\left[\partial_t\varphi - (\partial_\gamma n)M^{A\gamma} - n^\mu\partial_\mu\varphi^A + (\partial_\gamma n^\mu) F{}^A{}_\mu{}^\gamma\right](x)\,
\end{equation}
after previous execution of all related derivatives. 
The constraints are thus manifestly of at most first derivative order in the foliation parameter $t$, and the evolution equations of at most second derivative order in $t$, with respect to any chosen foliation. So there are, in particular, no Ostrogradsky ghosts \cite{Woodard}. The Helmholtz action giving rise to these canonical equations of motion is simply
\begin{equation}\label{Helmi}
S[\varphi,\pi,n,n^\alpha] = \int\mathrm dt\; \bigg\{ - H_t[\varphi,\pi,n,n^\alpha] + \int_\Sigma \mathrm d^3z \; \left(\pi_A \dot{\varphi}^A\right) (z)\bigg\} \,,
\end{equation} 
but, remarkably, can be expressed directly in terms of the functional $\mathcal{L}$ that follows from a solution of the gravitational closure equations. To see this, one uses the first Hamiltonian equation of motion in (\ref{Hamil}) above to express the derivative of the configuration variables with respect to the foliation parameter as
\begin{equation}\label{eq:shorthandK}
  \dot \varphi^A = n\, k^A[\varphi;\pi) + \left(\partial_\gamma n\right) M^{A\gamma}(\varphi) + \frac{\partial \widehat\varphi^A}{\partial g^\mathscr{A}}(\varphi) \left(\mathcal{L}_{\vec{n}} \widehat{g}(\varphi)\right)^{\mathscr{A}}\,.
\end{equation}
Upon insertion of this expression and the partially determined Hamiltonian  (\ref{eq:partialHam}) into the  Helmholtz action (\ref{Helmi}), one immediately observes that all terms but the ones coming from the local superhamiltonian drop out. Finally converting the $k^A$ back to the $\dot \varphi^A$ by use of the first Hamiltonian equation, one obtains an equivalent action
\begin{equation}
   S[\phi,N,N^\alpha] = \int\mathrm dt \int_\Sigma \mathrm d^3z \; \mathscr{L}_\text{\tiny geometry}[\phi,N,N^\alpha](t,z)\,
\end{equation}
where the capitalized quantities 
\begin{equation}
\phi(t,z):= \varphi_t(z)\,,\qquad N(t,z):=n_t(z)\,,\qquad N^\alpha(t,z):=n^\alpha_t(z)
\end{equation}
numerically precisely coincide with the configuration variables, the lapse and the shift, but are now all considered as spacetime quantities, rather than one-parameter families on the manifold $\Sigma$. In particular, functionals of the capitalized quantities may now include time derivatives, such as the Lagrangian density obtained by simple multiplication of the lapse $N$ with the solution $\mathcal{L}$ of the gravitational closure  equations,
\begin{eqnarray}\label{eq:scriL}
	\mathscr{L}_{\text{\tiny geometry}}[\phi,N,N^\alpha] = N \cdot \mathcal{L} \Big[\phi, \frac{1}{N} \Big((\dot{\phi}^A - \left(\partial_{\gamma}N\right) M^{A\gamma}(\phi) - \frac{\partial \widehat\varphi^A}{\partial g^\mathscr{A}}(\phi)  \left(\mathcal{L}_{\vec{N}} \widehat{g}(\phi)\right)^\mathscr{A}\Big)\Big)\,.
\end{eqnarray}
Indeed, it is quickly checked that varying the thus obtained total action
\begin{equation} 
S_\text{\tiny geometry}[\phi,N,N^\alpha] + S_\text{\tiny matter}[A;\phi,N,N^\alpha)
\end{equation}
with respect to the $\phi$, $N$ and $N^\alpha$ in a way that properly includes also time derivatives in the variations, yields a set of equations equivalent to the canonical gravitational evolution equations above.


\section{Examples: Matter on metric, bi-metric and higher-rank geometries}\label{sec_examples}
How truly simple it is now --- due to the new parameterization technology of any canonical geometry $g^\mathscr{A}$ in terms of non-tensorial configuration variables $\varphi^A$ --- to set up the gravitational closure equations for an admissible matter action on any tensorial background, is illustrated by the three hopefully instructive examples presented in this last section. The first one, in section \ref{sec_metricexample}, is a warm-up that starts from standard model matter, new only in that it uses non-tensorial configuration variables as the simplest illustration of how the latter are employed in practice. An illustration of how unexpectedly non-trivial the gravitational closure can turn out to be is then provided by the second example, which starts from an innocent-looking set of two scalar fields on a bi-metric background as the prescribed matter theory, for which the corresponding closure equations are set up in section \ref{sec_bimetricexample}. The last section \ref{sec_areametricexample} finally presents the gravitational closure equations for a gravity theory of some phenomenological interest, namely the one underpinning the most general birefringent linear electrodynamics. The seriously involved closure equations for this theory are solved perturbatively in \cite{SSS}.

\subsection{Gravitational closure of Klein-Gordon theory on a metric geometry} \label{sec_metricexample}
The arguably simplest canonically quantizable matter field theory on a metric background $(M,G)$, and thus one that necessarily satisfies the matter conditions imposed in section \ref{subsec_pdr}, is the Klein-Gordon action for a scalar field $\phi$,
\begin{equation}
   S_\textrm{\tiny matter}[\phi;G) = \int \mathrm  d^4x \sqrt{-(\det G_{\cdot\cdot})(x)} \left[G^{ab}(x) \partial_a \phi(x) \partial_b \phi(x) - m^2 \phi^2(x)\right]\,,
\end{equation}
whose principal tensor can be read off directly from the highest order derivative term of the associated field equations and has the components
\begin{equation}
P{}^{ij} = G^{ij}\,.
\end{equation}
All matter dynamics of the standard model of particle physics are constructed such that they feature this principal tensor. 
Thus the above Klein-Gordon theory, standard abelian and non-abelian gauge theory and indeed Dirac fields (the latter precisely because the Dirac algebra $\gamma^{(a} \gamma^{b)} = G^{ab}$ recovers again the same principal tensor) all produce the same triple
\begin{equation}
(M,G,G^{-1})\,,
\end{equation}
where the matter conditions (actually, in the metric case, the first matter condition already implies the second and the third) require the metric $G$ to have Lorentzian signature.

We now quickly rush through the steps described in this paper to set up the gravitational closure equations. First, the induced geometry is calculated to be
\begin{equation}
\mathbf{g}^{00} := G(\epsilon^0,\epsilon^0)\,,\qquad
\mathbf{g}^{0\alpha} := G(\epsilon^0, \epsilon^\alpha)\,, \qquad\
\mathbf{g}^{\alpha0} := G(\epsilon^\alpha, \epsilon^0)\,, \qquad\
\mathbf{g}^{\alpha\beta} := G(\epsilon^\alpha, \epsilon^\beta) \,.
\end{equation}
The associated frame conditions
\begin{equation}\label{metricframeconds}
\mathbf{p} = \mathbf{g}^{00} = 1  \qquad \text{and} \qquad \mathbf{p}^\alpha = \mathbf{g}^{\alpha0} = 0
\end{equation}
are obviously linear. Transition from the induced geometry to the corresponding 16 independent tensor field components $g$, $g^\alpha$ and $g^{\alpha\beta}$, and subsequent implementation of the frame constraints (\ref{metricframeconds}) together with the automatically linear symmetry constraints
\begin{equation}
g^{[\alpha\beta]}=0\qquad\textrm{ and }\qquad g^{[\alpha 0]} = 0 \,,
\end{equation}
removes $1+3+3+3$ tensor components. Thus, we are effectively only left with a symmetric tensor field $g^{\alpha\beta}$ that can be parametrized in terms of six configuration variables $\varphi^A$. Of the infinity of possible parametrizations, we choose the parametrization maps
\begin{equation}
\widehat{g}\,^{\alpha\beta}(\varphi) : = \I^{\alpha\beta}{}_{A} \, \varphi^{A} \qquad\textrm{ and }\qquad \widehat{\varphi}\,^A(g) := \I^A{}_{\alpha\beta}\, g^{\alpha\beta}\,,
\end{equation}
where the respective constant intertwining matrices 
need to satisfy the two conditions
\begin{equation}
\I^{\alpha\beta}{}_A \I^B{}_{\alpha\beta} = \delta^A_B \qquad\text{and}\qquad \I^{\gamma\delta}{}_A \I^A{}_{\alpha\beta} = \delta^{(\gamma}_\alpha \delta^{\delta)}_\beta
\end{equation}
in order to render the above pair a valid parametrization; a concrete choice \cite{Reiss} is
\begin{equation}
\I^{\alpha\beta}{}_A := \tfrac{1}{\sqrt{2}} \left[\begin{smallmatrix} \sqrt{2} & 0 & 0 & 0 & 0 & 0 \\ 0 & 1 & 0 & 0 & 0 & 0 \\ 0 & 0 & 1 & 0 & 0 & 0 \\ 0 & 1 & 0 & 0 & 0 & 0 \\ 0 & 0 & 0 & \sqrt{2} & 0 & 0 \\ 0 & 0 & 0 & 0 & 1 & 0 \\ 0 & 0 & 1 & 0 & 0 & 0 \\ 0 & 0 & 0 & 0 & 1 & 0 \\ 0 & 0 & 0 & 0 & 0 & \sqrt{2} \end{smallmatrix}\right]^{\alpha\beta}_{\,\,\, A} \quad \textrm{ and } \quad \I^{A}{}_{\alpha\beta} := \tfrac{1}{\sqrt{2}} \left[\begin{smallmatrix} \sqrt{2} & 0 & 0 & 0 & 0 & 0 & 0 & 0 & 0 \\ 0 & 1 & 0 & 1 & 0 & 0 & 0 & 0 & 0 \\ 0 & 0 & 1 & 0 & 0 & 0 & 1 & 0 & 0 \\ 0 & 0 & 0 & 0 & \sqrt{2} & 0 & 0 & 0 & 0 \\ 0 & 0 & 0 & 0 & 0 & 1 & 0 & 1 & 0 \\ 0 & 0 & 0 & 0 & 0 & 0 & 0 & 0 & \sqrt{2} \end{smallmatrix}\right]^A_{\,\,\alpha\beta}\,,
\end{equation}
which however is rarely needed explicitly. 
With this parametrization at hand, the input coefficients defining the specific gravitational closure equations for this case are quickly calculated to be given by
\begin{equation}
p^{\alpha\beta} = g^{\alpha\beta}\,,\qquad 
E^A{}_\mu = \varphi^A{}_{,\mu}\,,\qquad
F^A{}_\mu{}^\gamma = 2\, \I^A{}_{\mu\alpha}\, \I^{\gamma\alpha}{}_B \, \varphi^B \,,\qquad
M^{A\mu} = 0 \,.\\
\end{equation}
Solving the resulting gravitational closure equations yields \cite{Kuchar1974}, as the only non-vanishing dynamical potentials 
\begin{eqnarray}
C[\varphi] &=& -\frac{1}{2\kappa}\,\frac{1}{\sqrt{-\det\widehat{g}(\varphi)}} \big(R[\widehat g(\varphi)] - 2\,\Lambda\big)\,, \\
C_{AB}(\varphi) &=& \frac{1}{8\kappa}\,\frac{1}{\sqrt{-\det\widehat{g}(\varphi)}}\, \I^{\alpha\beta}{}_A \I^{\mu\nu}{}_B \big(\widehat{g}_{\alpha\mu}(\varphi) \widehat{g}_{\beta\nu}(\varphi) - \widehat{g}_{\alpha\beta}(\varphi) \widehat{g}_{\mu\nu}(\varphi)\big)\,,
\end{eqnarray}
where $R[g]$ denotes the Ricci curvature scalar built from an inverse three-dimensional metric $g$ and $\widehat{g}_{\alpha\beta}(\varphi)$ denotes the matrix inverse of $\widehat{g}^{\alpha\beta}(\varphi)$. But this is exactly the $3+1$ decomposition of the Einstein-Hilbert action
\begin{equation}
S_\textrm{\tiny geometry}[G] = \frac{1}{2\kappa} \int \mathrm  d^4x \sqrt{-\det G_{\cdot\cdot}(x)} \, \left[R(G(x),\partial G(x), \partial^2G(x)) + 2 \Lambda\right]
\end{equation}
with the gravitational constant $\kappa$ and cosmological constant $\Lambda$ having emerged as undetermined constants of integration.
Since nothing in our set-up has been designed to arrive at this result, the above is a successful test of the gravitational closure approach, as we know that the Einstein-Hilbert action is consistent with standard model matter. As indicated above, this result as such has been derived a long time ago by Kucha\u r and, indeed, our parametrization  of the canonical geometry $g^\mathscr{A}$ in terms of non-tensorial configuration variables $\varphi^A$ was a sledgehammer used to crack a nut, since the frame conditions were merely linear. But this will change dramatically already for the next, at first sight quite innocent-looking example of two free scalar fields coupled to two different metrics.

\subsection{Gravitational closure of two Klein-Gordon fields on a bi-metric geometry}\label{sec_bimetricexample}
A veritable surprise is in store when we consider the case of a bimetric geometry, featuring two (a priori not signature-restricted) metrics $G$ and $H$. In order to equip this geometry with specific kinematical meaning, we inject the physical information contained in the matter action
\begin{equation}\label{doubleKG}
S_\textrm{\tiny matter}[\phi,\psi;G,H) := \int \mathrm d^4x \left[\sqrt{-(\det G_{\cdot\cdot})(x)}\, G^{ab} \partial_a \phi(x) \partial_b \phi(x)  +  \sqrt{-(\det H_{\cdot\cdot})(x)}\, H^{ab}(x) \partial_a \psi(x) \partial_b \psi(x) \right]\,,
\end{equation}
for scalar fields $\phi$ and $\psi$, 
where additional terms giving rise to first and zeroth derivative order terms at the level of the associated equations of motion could be added at will, since they will not influence the principal tensor, which for this matter theory is \new{calculated to be the totally symmetrized product \cite{RRS} of the principal tensors of the two individual Klein-Gordon fields,
\begin{equation}\label{doublepoly}
P^{ijkl} = G^{(ij} H^{kl)}\,,
\end{equation}
which neatly illustrates the point made in section \ref{subsec_manyfields}, namely that a multitude of matter fields, and even a multitude of geometric tensors, still results --- as it must --- in one and only one principal tensor, which captures the 
information about the shared initial data surfaces, and thus a triple
\begin{equation}
  (M, \{G, H\}, P)\,
\end{equation}
form the matter field dynamics (\ref{doubleKG}).
Moreover, it is straightforward to see that the principal tensor $P$ provided by the above symmetrized product has the algebraic dual
\begin{equation}
  P^\#{}_{ijkl} = G_{(ij} H_{kl)}
\end{equation}
and that $P$ and $P^\#$ are both hyperbolic, as is required by the canonical quantizability of the matter action, if and only if both metrics $G$ and $H$ have Lorentzian signature.}

The induced geometry is constructed similarly to the case of one Lorentzian spacetime metric. However, there are now twice as many fields as there are two Lorentzian metrics $g^{\alpha\beta}$ and $h^{\alpha\beta}$, yielding
\begin{align}
\mathbf{g}^{00} := G(\epsilon^0,\epsilon^0)\,,\quad
\mathbf{g}^{0\alpha} := G(\epsilon^0, \epsilon^\alpha)\,, \quad\
\mathbf{g}^{\alpha0} := G(\epsilon^\alpha, \epsilon^0)\,, \quad\
\mathbf{g}^{\alpha\beta} := G(\epsilon^\alpha, \epsilon^\beta) \,,\nonumber\\
\mathbf{h}^{00} := H(\epsilon^0,\epsilon^0)\,,\quad
\mathbf{h}^{0\alpha} := H(\epsilon^0, \epsilon^\alpha)\,, \quad\
\mathbf{h}^{\alpha0} := H(\epsilon^\alpha, \epsilon^0)\,, \quad\
\mathbf{h}^{\alpha\beta} := H(\epsilon^\alpha, \epsilon^\beta) \,,
\end{align}
which satisfy, as always by construction of the employed frames, the frame  conditions, which in this case read
\begin{equation}
\mathbf{p} = \mathbf{g}^{00} \cdot \mathbf{h}^{00} = 1 \qquad\text{and}\qquad \mathbf{p}^\alpha = \frac{1}{2} (\mathbf{h}^{00} \, \mathbf{g}^{\alpha0} + \mathbf{g}^{00} \, \mathbf{h}^{\alpha0} ) = 0 \,.
\end{equation}
Transition to the corresponding canonical geometry $g, g^\alpha, g^{\alpha\beta}, h, h^\alpha, h^{\alpha\beta}$ requires to explicitly impose twelve symmetry conditions  
\begin{equation}
g^{[\alpha\beta]}=0\,,\qquad h^{[\alpha\beta]}=0\,,\qquad g^{[\alpha 0]} = 0\,, \qquad h^{[\alpha 0]} = 0 \,,
\end{equation}
in addition to the frame conditions above, which reduce to requiring that
\begin{equation}
h^{00} = \frac{1}{g^{00}} \qquad \textrm{ and } \qquad h^{\alpha0} = - \frac{1}{(g^{00})^2} \, g^{\alpha 0} \,.
\end{equation}
One thus finds that only 16 of the 32 components of the canonical geometry are independent. Unlike in the mono-metric case, however, the parametrization of the canonical geometry in terms of non-tensorial configuration variables, as developed in this paper, is now seriously needed, since the frame conditions are non-linear. 
Since it helps to group the relevant expressions, it is convenient to introduce the card game notation
\begin{equation}
\varphi^A =: (\lp, \ll^1, \ll^2, \ll^3, \lll^1, \dots, \lll^6, \dots, \llll^1, \dots, \llll^6)^A
\end{equation}
for $A=1,\dots,16$, with the various groups of configuration variables mirroring the corresponding groups of tensors making up the canonical geometry, which we choose to parametrize as
\begin{equation}
\widehat{g}(\varphi) := \lp \,,\qquad
\widehat{g}^{\,\alpha}(\varphi) := \I^\alpha{}_a \, \ll^a \\, \qquad 
\widehat{g}^{\,\alpha\beta}(\varphi) := \I^{\alpha\beta}{}_A \, \lll^A \,,\qquad
\widehat{h}^{\alpha\beta}(\varphi) := \I^{\alpha\beta}{}_A \, \llll^A\,,
\end{equation}
where a lowercase Latin index $a$ ranges over $1,2,3$, while an uppercase Latin index $A$ ranges over $1,\dots, 6$. The constant intertwining matrices $\I^{\alpha\beta}{}_A$ and $\I{}^A{}_{\alpha\beta}$ are as in the previous example of a mono-metric geometry, while 
\begin{equation}
\I^\alpha{}_a := \left[\begin{smallmatrix}
1 & 0 & 0 \\ 0 & 1 & 0 \\ 0 & 0 & 1
\end{smallmatrix}\right]^\alpha_{\,\,\,a} \qquad \textrm{ and } \qquad \I^a{}_\alpha := \left[\begin{smallmatrix}
1 & 0 & 0 \\ 0 & 1 & 0 \\ 0 & 0 & 1
\end{smallmatrix}\right]^a_{\,\,\,\alpha}\,,
\end{equation}
and thus satisfy
$\I^\alpha{}_b \, \I^a{}_\alpha = \delta^a_b$ and $\I^\alpha{}_a \, \I^a{}_\beta = \delta^\alpha_\beta$. 
The input coefficients are then straightforwardly calculated. Whenever it is convenient to keep terms and notation short and clear, the split of the configuration variables devised above will also be used in the expressions for the input coefficients. The input coefficients are 
$$
p^{\alpha\beta} = \frac{1}{6\lp} \, \I^{\alpha\beta}{}_A\,\left( \lll^A + \lp^2 \llll^A\right) - \frac{2}{3 (\lp)^2}\, \I^\alpha{}_a \I^\beta{}_b \, \ll^a \ll^b\,,
$$
$$E^A{}_\mu = \varphi^A{}_{,\mu} \,,$$
$$F^{\overline{\cdot}}{}_\mu{}^\gamma = 0 \,,\quad
F^{\overline{\overline{a}}}{}_\mu{}^\gamma = \I^a{}_\mu \I^\gamma{}_b \, \ll^b \,,\quad
F^{\overline{\overline{\overline{A}}}}{}_\mu{}^\gamma = 2\, \I^A{}_{\mu\alpha} \I^{\gamma\alpha}{}_B \, \lll^B \,,\quad
F^{\overline{\overline{\overline{\overline{A}}}}}{}_\mu{}^\gamma = 2\, \I^A{}_{\mu\alpha} \I^{\gamma\alpha}{}_B \, \llll^B \,,$$
$$
M^{\overline{\cdot}\,\gamma} = - 2\,\I^\gamma{}_a \, \ll^a \,,
$$
$$
M^{\overline{\overline{a}}\,\gamma} = \frac{1}{2}\, \I^a\,_\alpha \I^{\alpha\gamma}_A \left( (\lp)^2 \llll^A - \lll^A \right) - \frac{2}{\lp} \, \I^\gamma{}_b \, \ll^a \, \ll^b \,,\\
$$
$$
M^{\overline{\overline{\overline{A}}}\,\gamma} = \frac{1}{\lp} \, \I^A{}_{\alpha\beta} \I^\alpha{}_a \I^{\beta\gamma}{}_B\,\ll^a \left(\lll^B +(\lp)^2 \, \llll^B \right) - \frac{4}{(\lp)^2}\, \I^A{}_{\alpha\beta} \I^\alpha{}_a \I^\beta{}_b \I^\gamma{}_c \, \ll^a \, \ll^b \, \ll^c \,,\\
$$
\begin{equation}
M^{\overline{\overline{\overline{\overline{A}}}}\,\gamma} = -\frac{1}{(\lp)^3} \I^A{}_{\alpha\beta} \I^\alpha{}_a \I^{\beta\gamma}{}_B \,\ll^a \left(\lll^B + (\lp)^2  \llll^B\right) + \frac{4}{(\lp)^4}\, \I^A{}_{\alpha\beta} \I^\alpha{}_a \I^\beta{}_b \I^\gamma{}_c \, \ll^a \, \ll^b \, \ll^c \,.\label{bimetricinput}
\end{equation}
With these input coefficients, the gravitational closure equations can be set up. 

It is evident that the case of a bi-metric spacetime does not decompose into two separate metric sectors, as is often intuitively assumed, since then the fact that one shared principal tensor is required in order to allow for a common evolution from common initial data surfaces would not be taken into account. Finding the most general Lagrangian for bi-metric gravity is as complicated as solving the gravitational closure equations specialized to the input coefficients (\ref{bimetricinput}). Their explicit solution is an open problem to be solved if one proposes such a theory.  The linearized gravitational field equations, however, have already been obtained \cite{BSW}.

\subsection{Gravitational closure of general linear electrodynamics}\label{sec_areametricexample}
We finally set up the gravitational closure equations for the refinement of Maxwell theory \new{that equips an abelian gauge covector field $A$ with the dynamics
\begin{equation}\label{areametricED}
  S_\textrm{\tiny matter}[A;G) =  \int \mathrm d^4x (\epsilon_{pqrs} G^{pqrs}(x))^{-1} G^{abcd}(x) F_{ab}(x) F_{cd}(x) \qquad \textrm{for } F_{ab} := \partial_a A_b - \partial_b A_a
\end{equation}
on an orientable four-dimensional area metric manifold \cite{SW}, which carries the canonical top form density $\epsilon$ with $\epsilon_{0123}=1$ and a fourth rank contravariant tensor field $G$ featuring the algebraic symmetries $G^{abcd}=G^{cdab}$ and $G^{abcd} = G^{[ab][cd]}$ and satisfying $\epsilon_{abcd} G^{abcd}\neq 0$ everywhere.} 
The principal tensor of this theory has been calculated first by Rubilar \cite{Rubilar, Rubilaretal} and takes the form
\begin{equation}
  P^{ijkl} = -\tfrac{1}{24} ( \tfrac{1}{24}\epsilon_{abcd} G^{abcd})^{-2} \epsilon_{mnpq} \epsilon_{rstu} G^{mnr(i} G^{j|ps|k} G^{l)qtu}\,,
\end{equation}
whose non-polynomial dependence of the geometric tensor $G$ presents a technically particularly involved kinematical structure. The requirement that the above general linear electrodynamics satisfy the three matter conditions requires that $G$ lie in one of seven (out of a total 23) algebraic classes  \cite{SW}. The induced geometry features fields with antisymmetric index pairs, which we can dualize using the volume form density on $\Sigma$, arriving at the set
\begin{eqnarray}
\mathbf{\overline{g}}^{\alpha\beta} &:=& - G({\epsilon^0,\epsilon^\alpha,\epsilon^0,\epsilon^\beta})\,, \\
\mathbf{\overline{\overline{g}}}_{\alpha\beta} &:=& \frac{1}{4} \frac{1}{\det \mathbf{\overline{g}^{\cdot\cdot}}}\, \epsilon_{\alpha\mu\nu}\epsilon_{\beta\rho\sigma}  G({\epsilon^\mu,\epsilon^\nu,\epsilon^\rho,\epsilon^\sigma})\,, \\
\mathbf{\overline{\overline{\overline{g}}}}_{\alpha\beta} &:=& (\mathbf{\overline{g}}^{-1})_{\alpha\mu} \left( \frac{1}{2} \frac{1}{\sqrt{\det \mathbf{\overline{g}^{\cdot\cdot}}}}\,\epsilon_{\beta\kappa\lambda} G({\epsilon^0,\epsilon^\mu,\epsilon^\kappa,\epsilon^\lambda}) - \delta^\mu_\beta\right)\,.
\end{eqnarray}
The frame conditions for the employed frames, expressed in terms of the induced fields, are
\begin{eqnarray*}
	\mathbf{\overline{g}}^{\alpha\beta} \mathbf{\overline{\overline{\overline{g}}}}_{\alpha\beta} = 0 \qquad \textrm{ and } \qquad \mathbf{\overline{\overline{\overline{g}}}}_{[\alpha\beta]} = 0\,.
\end{eqnarray*}

Transition to the corresponding canonical geometry $\G^{\alpha\beta}, \GG_{\alpha\beta}, \GGG_{\alpha\beta}$ thus requires to explicitly enforce these four conditions, together with the remaining symmetry conditions by requiring that
\begin{equation}
\G^{\alpha\beta} \GGG_{\alpha\beta} = 0\,,\qquad \GGG_{[\alpha\beta]} = 0\,,\qquad  \G^{[\alpha\beta]} = 0\,,\qquad  \GG_{[\alpha\beta]} = 0\,,
\end{equation}
reducing the a priori $27$ independent entries of the tensor fields that make up the canonical geometry by 10. In order to account for these conditions, we thus need to choose 17 unconstrained configuration variables. It is convenient to denote them by
\begin{equation}
\varphi^A := (\lp^1, \dots, \lp^6, \ll^1, \dots, \ll^6, \lll^1, \dots, \lll^5)
\end{equation}
and to construct the parametrization maps ($a,b,c =1\dots6$ and $m,n =1\dots 5$)
\begin{equation}
\widehat{\overline g}{}^{\alpha\beta}(\varphi) := \I^{\alpha\beta}{}_a \lp^a\,,\qquad
\widehat{\overline{\overline g}}{}_{\alpha\beta}(\varphi) := \I^a{}_{\alpha\beta} \Delta_{ab} \ll^b \,,\qquad
\widehat{\overline{\overline{\overline g}}}{}_{\alpha\beta}(\varphi) := \I^a{}_{\alpha\beta} \left(\delta_a^b - \frac{n_a \lp^b}{n_c\lp^c} \right) \epsilon_{(m)b} \lll^m \,,
\end{equation}
where $\Delta_{ab}$ are the constant components of the standard inner product on $\mathbb{R}^6$, and $t^a, e^{(1)a}\,, \dots,\, e^{(5)a}$ are
the components of constant orthonormal basis vectors chosen such that $\I^a{}_{\alpha\beta} \Delta_{ab} t^b$ is a positive definite matrix. 
Note that $n_a := \Delta_{ab} t^b\,, \epsilon_{(1)a} := \Delta_{ab} e^{(1)b}\,, \dots,\,   \epsilon_{(5)a} := \Delta_{ab} e^{(5)b}$ is then the dual basis.
Conversely, extraction of the configuration variables $\varphi^A$ from the tensor fields $g^\mathscr{A}$ constituting the canonical geometry is achieved by the maps
\begin{equation}
\widehat{\lp}{}^a(g) := \I^a{}_{\alpha\beta} \overline{g}{}^{\alpha\beta}\,,\qquad
\widehat{\ll}{}^a(g) := \Delta^{ab} \I^{\alpha\beta}_b \overline{\overline{g}}_{\alpha\beta} \,, \qquad
\widehat{\lll}{}^m(g) := \I^{\alpha\beta}{}_a e^{(m)a} \overline{\overline{\overline g}}{}_{\alpha\beta}\,,
\end{equation}
which indeed recover precisely the configuration variables employed in the parametrization, as one readily checks.
It is clear by construction that the three maps $\widehat{\overline{g}}$, $\widehat{\overline{\overline{g}}}$, $\widehat{\overline{\overline{\overline{g}}}}$ produce symmetric tensor fields, so that the last three conditions above are obviously satisfied, while
\begin{equation}
\widehat{\overline{g}}{}^{\alpha\beta}(\varphi) \widehat{\overline{\overline{\overline{g}}}}{}_{\alpha\beta}(\varphi) = \I^{\alpha\beta}{}_a \I^b{}_{\alpha\beta}\left(\delta_b^c - \frac{n_b \lp^c}{n_d \lp^d}\right) \epsilon_{(m)c} \lp^a \lll^m = \epsilon_{(m)a} \lp^a \lll^m - \epsilon_{(m)a} \lp^a \lll^m = 0 \,,
\end{equation}
shows that also the first condition above is satisfied. The intertwiners associated with this parametrization are then readily calculated as 
\begin{equation}
	\frac{\partial\widehat{\overline{g}}{}^{\alpha\beta}}{\partial\lp^a} = \I^{\alpha\beta}{}_a \,, \qquad
	\frac{\partial\widehat{\overline{\overline{g}}}{}_{\alpha\beta}}{\partial\ll^a} = \Delta_{ab} \I^b{}_{\alpha\beta} \,, \qquad
	\frac{\partial\widehat{\overline{\overline{\overline{g}}}}{}_{\alpha\beta}}{\partial\lll^m} = \I^a{}_{\alpha\beta} \left( \delta^b_a - \frac{n_a \lp^b}{n_c \lp^c} \right) \epsilon_{(m)b}\,, 
\end{equation}
\begin{equation}
	\frac{\partial\widehat{\overline{\overline{\overline{g}}}}{}_{\alpha\beta}}{\partial\lp^a} = \I^b{}_{\alpha\beta}n_b \frac{n_a \lp^c \epsilon_{(m)c} \lll^{m}}{(n_d \lp^d)^2} - \I^b{}_{\alpha\beta} n_b \frac{\epsilon_{(m)a} \lll^m}{n_c \lp^c}\,, 
\end{equation}
\begin{equation}
	\frac{\partial\widehat{\lp}{}^a}{\partial \overline{g}{}^{\alpha\beta}} = \I^a{}_{\alpha\beta} \,, \qquad
	\frac{\partial\widehat{\ll}{}^a}{\partial \overline{\overline{g}}{}_{\alpha\beta}} = \Delta^{ab} \I^{\alpha\beta}{}_b \,, \qquad
	\frac{\partial\widehat{\lll}{}^m}{\partial \overline{\overline{\overline{g}}}{}_{\alpha\beta}} = \I^{\alpha\beta}{}_a e^{(m)a} \,.
    \end{equation}
The input coefficients for the gravitational closure equations are therefore
$$
  p^{\alpha\beta} = \frac{1}{6} \left( \g^{\alpha\gamma} \g^{\beta\delta} \gg_{\gamma\delta} - \g^{\alpha\beta} \g^{\gamma\delta} \gg_{\gamma\delta} - 2 \g^{\alpha\beta} \g^{\delta\mu} \g^{\gamma\nu} \ggg_{\mu\gamma} \ggg_{\nu\delta} + 3 \g^{\gamma\delta} \g^{\alpha\mu} \g^{\beta\nu} \ggg_{\mu\gamma} \ggg_{\nu\delta} \right)\,,
$$ 
$$
	\overline{E}{}^a{}_\mu = \lp^a{}_{,\mu} \,,\qquad  
	\overline{\overline{E}}{}^a{}_\mu = \ll^a{}_{,\mu} \,,\qquad 	
	\overline{\overline{\overline{E}}}{}^m{}_\mu = \lll^a{}_{,\mu} \,,
$$
$$
	\overline{F}{}^a{}_\mu{}^\gamma = 2 \I^a{}_{\gamma\alpha} \I^{\mu\alpha}{}_{b} \lp^{b}\,, \qquad
	\overline{\overline{F}}{}^a{}_\mu{}^\gamma = -2 \Delta^{ab} \Delta_{cd} \I^{\gamma\alpha}{}_b \I^c{}_{\mu\alpha} \, \ll^{d}\,, \qquad	
	\overline{\overline{\overline{F}}}{}^m{}_\mu{}^\gamma = - 2  \frac{\partial\widehat{\lll}{}^m}{\partial \overline{\overline{\overline{g}}}{}_{\gamma\alpha}} \frac{\partial\widehat{\overline{\overline{\overline{g}}}}{}_{\mu\alpha}}{\partial\lll^n}  \,\lll^{n}\,,
$$
$$
\overline{M}{}^{a\gamma}  =  2 (\det\widehat{\G}{}^{\cdot\cdot}(\varphi))^{1/2} \, \I^a{}_{\alpha\beta} \I^{\nu(\alpha|}{}_b \epsilon^{|\beta)\mu\gamma} \frac{\partial\widehat{\overline{\overline{\overline{g}}}}{}_{\mu\nu}}{\partial\lll^m}(\varphi) \,\lp^{b} \lll^{m} \,,
$$
$$
\overline{\overline{M}}{}^{a\gamma}  =  6  (\det\widehat{\G}{}^{\cdot\cdot}(\varphi))^{-1/2}  \, \epsilon_{\alpha\mu\nu} \Delta^{ab} \I^{\alpha\beta}{}_b \I^{\lambda\nu}{}_c p^{\mu\gamma}(g(\varphi)) \frac{\partial\widehat{\overline{\overline{\overline{g}}}}{}_{\beta\lambda}}{\partial\lll^m}(\varphi) \, \lp^c \lll^m  \,,
$$
\begin{equation}
\overline{\overline{\overline{M}}}{}^{m\gamma} = -   (\det\widehat{\G}{}^{\cdot\cdot}(\varphi))^{1/2} \epsilon^{\mu\nu\gamma} \, \left(\widehat{\G}^{-1}\right){}_{\mu\alpha} \frac{\partial\widehat{\lll}{}^m}{\partial \overline{\overline{\overline{g}}}{}_{\alpha\beta}}(\varphi)  \left( \I^{\kappa\lambda}{}_b \frac{\partial\widehat{\overline{\overline{\overline{g}}}}{}_{\beta\lambda}}{\partial\lll^n}(\varphi) \frac{\partial\widehat{\overline{\overline{\overline{g}}}}{}_{\kappa\nu}}{\partial\lll^l}(\varphi) \lp^{b} \lll^{n} \lll^{l} + \I^b{}_{\beta\nu} \Delta_{bc} \ll^c \right)\,.\label{areametricinput}
\end{equation}
The corresponding gravitational closure equations differ significantly from those proposed for this case in \cite{GSWW}, because now the non-linear frame conditions are already taken care of by our use of non-tensorial configuration degrees of freedom, while previously they had to be added by hand and thus made the problem of solving the equations prohibitively difficult. 

An exact solution of the gravitational closure equations for the general linear electrodynamics (\ref{areametricED}) is hard to obtain, due to the complicated input coefficients (\ref{areametricinput}). But even if exact solutions of the closure equations were obtained, they would just lead to exact field equations for an area metric, which in turn one typically would have to solve either by imposing some symmetry assumption or by resorting to perturbation theory --- as is already the case for the standard Einstein field equations for a Lorentzian metric. For this reason, one may equally well aim at ultimately linearized or symmetrized gravitational field equations already at the level of the closure equations. Carefully taking into account how far truncated partial differential equations may be evaluated and under which circumstances symmetry conditions may be inserted already at the level of the action, meanwhile these strategies have been implemented successfully, leading to linearized \cite{SSS} and cosmological \cite{DFS} gravitational field equations for an area metric.

\section{Conclusions} \label{sec_conclusions}
We showed how to gravitationally close a given set of matter field equations, in the sense of providing equations of motion for the background geometry on which the matter dynamics have been formulated in the first place. Practically, this is done by following the concrete calculational sequence
\begin{center}
 {\it matter equations  $\rightarrow$ input coefficients $\rightarrow$ output coefficients $\rightarrow$ gravity equations.}
\end{center} 
The first step, at its core, is a straightforward standard calculation in the theory of partial differential equations, namely the calculation of the principal tensor of the matter field equations one starts from. It is then easy to identify the canonical geometry and to parametrize the latter in terms of non-tensorial configuration variables such that generically non-linear frame conditions are automatically captured and thus need not be worried about anymore in the remaining course of the treatment. If the matter dynamics are canonically quantizable, the previously calculated principal tensor features all the properties needed to calculate the input coefficients that are required to set up the gravitational closure equations, \new{although the actually required conditions of section \ref{subsec_pdr} on the matter dynamics are generically weaker than their canonical quantizability.}
The second step then consists in solving the gravitational closure equations for the output coefficients, yielding a gravitational Hamiltonian subject to the conditions of section \ref{subsec_HKT}. Depending on the complexity incurred by the specific input coefficients at hand, one may be able to find their general local solution, have to resort to perturbative techniques or employ a symmetry reduction in order to extract physical predictions. 
The third step is again straightforward, as it merely consists of employing the output coefficients to compose the gravitational action, whose variation with respect to the configuration variables then yields one side of the thus defined gravitational field equations. The other side is of course provided by the same variation, but applied to the matter action from which the entire construction started. 

There are only a few routes by which one can escape the gravitational closure mechanism when presented with a matter theory coupled to some geometry. One is to introduce, in addition to the geometry employed in the matter dynamics, additional gravitational degrees of freedom to which none of the matter fields couple directly; this allows for arbitrary modifications to the gravitational dynamics, 
and thus comes at the cost of needing an infinite number of experiments to determine the constants of the theory before it becomes numerically predictive. \new{The other circumvention would be to drop matter conditions for at least some of the matter that inhabits the universe one wishes to model. This would certainly exclude canonically quantizable matter, but also, depending on how many of the three matter conditions are violated, would prevent a consistent classical notion of massive particles in the best case, or additionally an observer-independent notion of positive energy, or, in the worst case, additionally massless particles. We believe that this cascade of problems, incurred when dropping our matter assumptions underlines the physical soundness of the latter, and thus that of the gravitational closure procedure built on it.}



Fundamentally, the ability to perform the gravitational closure of canonically quantizable matter dynamics allows us to inject our current and future knowledge about matter directly into the construction of gravity theories.
Additional constraints, such as the absence of ghosts, can and should be employed to further reduce the linear solution space of the gravitational closure equations. However, it is typically the specific gravitational closure equations as they follow from concrete matter dynamics --- and not sweeping theoretical constraints --- that effectively reduce the spectrum of possible gravity theories, such that, at best, only a finite number of constants are left to be determined by observation. Indeed, even decisive generic requirements, such as diffeomorphism invariance or ghost-freedom, do generally not achieve that.

Phenomenologically, one can now ask questions that hitherto were not systematically accessible, since they require bridging the gap between a hypothesis about matter and the resulting gravitational implications. For instance, a systematic exploration of the simple question whether there is any evidence for birefringence of light in vacuo, compels one to forsake the assumption of a metric background geometry in a favor of a refinement  \cite{Hehlbook,biref1,biref2,biref3,biref4} that can be written by  a fourth rank tensor $G$, such that Maxwell's action is refined to the general linear electrodynamics, whose gravitational closure equations we derived in section \ref{sec_areametricexample}. The refined Maxwell theory is canonically quantizable \cite{RS,Pfeifer,GST} and thus provides valid input coefficients for the pertinent gravitational closure equations. The temptation to discard such a refinement a priori 
is quite delusive. For even if coarse geometric optics effects are undetectable, the above action still predicts accumulative modifications for the way electromagnetic field energy is transported \cite{WittePhD}. These result in a potentially measurable modification of Etherington's distance duality relation \cite{SW1} already in a weak gravitational field that admits birefringence \cite{SSS}, and which may also address otherwise inexplicable magnification anomalies \cite{magnificationanomaly}. \\[3pt]
\indent Based on the results of the present paper, we believe that the construction of gravity theories must consider the dynamics of all matter fields that will populate a spacetime right from the start. The gravitational closure  equations enable one to put this insight to immediate practical use. Either as a complete consistency check for an existing gravity theory, or for its derivation.

\section*{Acknowledgments} 
The authors thank Marcus Werner, Shinji Mukhoyama, Antonio de Felice, Jonas Schneider and Felix Finster for most valuable comments and discussions. NS gratefully acknowledges support by the Studienstiftung des deutschen Volkes and FPS thanks the Yukawa Institute for Theoretical Physics for the invitation to a two-month research collaboration building on the results of this paper.

\bibliography{references}{}

\newpage

\begin{turnpage}
\begin{table}\label{pagemaster}
\vspace{-2cm}
\begin{small}
\centering{\bf GRAVITATIONAL CLOSURE  EQUATIONS} 
\begin{align*}
& \qquad\qquad \textrm{\straight{Input coefficients} \qquad\,\,\, \textrm{defined by}}  \qquad   N^\mu \straight{E^A{}_\mu}[\varphi] - \partial_\gamma N^\mu \straight{F^{A}{}_\mu{}^\gamma}(\varphi)  := \frac{\partial \widehat \varphi^A}{\partial g^\mathscr{A}} \left(\mathcal{L}_{\vec{N}} \widehat g(\varphi)\right)^\mathscr{A}\quad\textrm{ and } \quad \tilted{M^{A}{}^{\gamma}}(\varphi) := \frac{\partial \widehat\varphi^A}{\partial g^\mathscr{A}}(\widehat g(\varphi)) \, e_0^a(t,\sigma) \frac{\partial\mathbf{g}^\mathscr{A}}{\partial\partial_\gamma X^a}(t,\sigma)\\[3pt]
& \qquad\qquad \textrm{\coeffs{Output coefficients} \quad\quad defining} \,\,\,\qquad \,\,\,  \mathcal{L}[\varphi;K) :=  \sum_{N=0}^\infty \coeffs{C_{A_1\dots A_N}}[\varphi] K^{A_1} \cdots K^{A_N}
\end{align*}
$ $ \qquad \quad\,\, The finite upper limit $\max$ is to be determined,  individually for each of the sums below, according to the criteria in section \ref{subsec_towardsconstruction}.
$ $\\[12pt]
\centering{\bf The seven individual equations}
\begin{align*}
&\qquad\qquad \col{1} \quad 0 = -\coeffs{C} \delta^\gamma_\mu + \sum_{K=0}^{\textrm{\tiny max}} (K+1) \CC{}{A}{\gamma\alpha_1\dots\alpha_K} \left(\straight{E^A{}_{\mu,\alpha_1\dots\alpha_K}} + \straight{F^A{}_\mu{}^{\alpha_{K+1}}{}_{,\alpha_1\dots\alpha_{K+1}}} \right) - \sum_{K=0}^{\textrm{\tiny max}} (K+1) \CC{}{A}{(\alpha_1\dots\alpha_K|} \straight{F^A{}_\mu{}^{|\gamma)}{}_{,\alpha_1\dots\alpha_K}} \\[0pt]
&\qquad\qquad\col{2} \quad 0 = -\Coeff{A} \left( \straight{E^A{}_{\mu:B}{}^{\gamma}} + \straight{F^A{}_{\mu}{}^\gamma{}_{:B}} \right) + \sum_{K=0}^{\textrm{\tiny max}} (K+1) \CC{B}{A}{\gamma\alpha_1\dots\alpha_K} \left(\straight{E^A{}_{\mu,\alpha_1\dots\alpha_K}} + \straight{F^A{}_\mu{}^{\alpha_{K+1}}{}_{,\alpha_1\dots\alpha_{K+1}}} \right) - \sum_{K=0}^{\textrm{\tiny max}} (K+1) \CC{B}{A}{(\alpha_1\dots\alpha_K|} \straight{F^A{}_\mu{}^{|\gamma)}{}_{,\alpha_1\dots\alpha_K}} \\[0pt]
&\qquad\qquad\col{3} \quad 0 = 2\left(\mathrm{deg} P-1\right) \init{p^{(\mu|\rho}}\Coeff{AB} \straight{F^A{}_\rho{}^{|\nu)}} + \sum_{K=0}^{\textrm{\tiny max}} (K+1)\CC{B}{A}{\alpha_1\dots\alpha_K(\mu|} \tilted{M^{A|\nu)}{}_{,\alpha_1\dots\alpha_K}} - \sum_{K=0}^{\textrm{\tiny max}} (-1)^K \binom{K+2}{K} \left( \partial^K_{\alpha_1\dots\alpha_K} \CC{}{B}{\alpha_1\dots\alpha_K\mu\nu} \right)
\\[0pt]
&\qquad\qquad\col{4} \quad 0 = 2 \left(\mathrm{deg} P-1\right) \Coeff{AB} \left( \init{p^{\mu\nu}} \straight{E^A{}_\nu} - \init{p^{\mu\nu}{}_{,\gamma}} \straight{F^A{}_\nu{}^{\gamma}}\right) - \Coeff{A} \tilted{M^{A\mu}{}_{:B}} - \sum_{K=0}^{\textrm{\tiny max}} \CC{B}{A}{\alpha_1\dots\alpha_K} \tilted{M^{A\mu}{}_{,\alpha_1\dots\alpha_K}} - \sum_{K=0}^{\textrm{\tiny max}}  (-1)^K (K+1) \left( \partial^K_{\alpha_1\dots\alpha_K} \CC{}{B}{\alpha_1\dots\alpha_K\mu} \right) \\[0pt]
&\qquad\qquad\col{5} \quad 0 = 2\partial_\mu\left(\Coeff{A}\tilted{M^{A[\mu|}{}_{:B} M^{B|\gamma]}}\right) - 2 \left(\mathrm{deg} P-1\right) \init{p^{\rho\gamma}}\bigg[ \Coeff{A} \straight{E^A{}_\rho} + \partial_\mu\left( \Coeff{A} \straight{F^A{}_\rho{}^\mu}\right)  \bigg] + \sum_{K=0}^{\textrm{\tiny max}} \CC{}{A}{\alpha_1\dots\alpha_K} \tilted{M^{A\gamma}{}_{,\alpha_1\dots\alpha_K}} + \\[0pt]
&\qquad\qquad\qquad\qquad  +\sum_{K=0}^{\textrm{\tiny max}} \sum_{J=0}^K (-1)^J \binom{K}{J} (J+1) \partial^J_{\alpha_1\dots\alpha_J} \left( \CC{}{A}{\beta_1\dots\beta_{K-J}(\alpha_1\dots\alpha_J|} \tilted{M^{A|\gamma)}{}_{,\beta_1\dots\beta_{K-J}}} \right) \\ 
&\qquad\qquad\col{6} \quad 0 = 6 \left(\mathrm{deg} P-1\right) \Coeff{AB_1 B_2} \left( \init{p^{\mu\nu}} \straight{E^A{}_\nu} - \init{p^{\mu\nu}{}_{,\gamma}}\straight{F^A{}_\nu{}^{\gamma}}\right)  - 4 \Coeff{A(B_1} \tilted{M^{A\mu}{}_{:B_2)}} - 2 \CC{B_1 B_2}{A}{} \tilted{M^{A\mu}} - 2 \CC{B_1 B_2}{A}{\alpha} \tilted{M^{A\mu}{}_{,\alpha}} - 2 \CC{B_1 B_2}{A}{\alpha\beta} \tilted{M^{A\mu}{}_{,\alpha\beta}} \\[0pt]
&\qquad\qquad\qquad\qquad   - \CC{B_2}{B_1}{\mu} - \sum_{K=0}^{\textrm{\tiny max}} (-1)^K (K+1) \left( \partial^K_{\alpha_1\dots\alpha_K} \CC{B_1}{B_2}{\mu\alpha_1\dots\alpha_K} \right)  \\[0pt]
&\qquad\qquad\col{7} \quad 0  =  \sum_{K=2}^{\textrm{\tiny max}} \sum_{J=2}^{K} (-1)^J \binom{K}{J} (J-1) \partial^{J+1}_{\gamma\alpha_1\dots\alpha_{J}} \left( \CC{}{A}{\beta_1\dots\beta_{K-J} (\alpha_1 \dots \alpha_J|} \tilted{M^{A|\gamma)}{}_{,\beta_1\dots\beta_{K-J}}} \right)
\end{align*}
\end{small}
\end{table}
\end{turnpage} 

\begin{turnpage}
\begin{table}\label{pagecov}
\vspace{-2cm}
\begin{small}
\centering{\bf The fourteen sequences of equations for $N\geq 2$} 
\begin{align*}
&\qquad \qquad\col{8${}_N$} \quad 0 = \sum_{K=0}^{\textrm{\tiny max}} \binom{K+N}{K} \left[ \CC{}{A}{\beta_1\dots\beta_N \alpha_1\dots \alpha_K} \left( \straight{E^A{}_{\mu,\alpha_1\dots\alpha_K}} + \straight{F^A{}_\mu{}^{\alpha_{K+1}}{}_{,\alpha_1\dots\alpha_{K+1}}} \right) - \CC{}{A}{(\beta_1\dots\beta_N \alpha_1\dots \alpha_{K-1}|} \straight{F^A{}_\mu{}^{|\alpha_K)}{}_{,\alpha_1\dots \alpha_K}} \right]\\
&\qquad \qquad\col{9${}_N$} \quad 0 = \sum_{K=0}^{\textrm{\tiny max}} \binom{K+N}{K} \left[ \CC{B}{A}{\beta_1\dots\beta_N \alpha_1\dots \alpha_K} \left( \straight{E^A{}_{\mu,\alpha_1\dots\alpha_K}} + \straight{F^A{}_\mu{}^{\alpha_{K+1}}{}_{,\alpha_1\dots\alpha_{K+1}}} \right) - \CC{B}{A}{(\beta_1\dots\beta_N \alpha_1\dots \alpha_{K-1}|} \straight{F^A{}_\mu{}^{|\alpha_K)}{}_{,\alpha_1\dots \alpha_K}} \right]\\
& \qquad \qquad\col{10${}_N$} \quad 0 = - \Coeff{B_1\dots B_N} \delta^\gamma_\mu - N \Coeff{A(B_1\dots B_{N-1}} \straight{F^A{}_{|\mu|}{}^\gamma{}_{:B_N)}} - \CC{B_1\dots B_N}{A}{} \straight{F^A{}_\mu{}^\gamma} + \CC{B_1\dots B_N}{A}{\gamma} \straight{E^A{}_\mu} - \CC{B_1\dots B_N}{A}{\alpha} \straight{F^A{}_\mu{}^\gamma{}_{,\alpha}}  \\[0pt]
&\qquad\qquad\qquad\qquad \qquad + 2 \,\CC{B_1\dots B_N}{A}{\alpha\gamma} \straight{E^A{}_{\mu,\alpha}} - \CC{B_1\dots B_N}{A}{\alpha\beta} \straight{F^A{}_\mu{}^\gamma{}_{,\alpha\beta}} \\[0pt]
&\qquad\qquad\col{11${}_N$} \quad 0 = \CC{B_1\dots B_N}{A}{\beta_1\beta_2} \straight{E^A{}_\mu} - \CC{B_1\dots B_N}{A}{(\beta_1|} \straight{F^A{}_\mu{}^{|\beta_2)}} - 2 \CC{B_1\dots B_N}{A}{\alpha(\beta_1|} \straight{F^A{}_\mu{}^{|\beta_2)}{}_{,\alpha}} \\[\tunedvs]
&\qquad\qquad\col{12${}_N$} \quad 0 = \CC{B_1\dots B_N}{A}{(\alpha\beta|} \straight{F^A{}_\mu{}^{|\gamma)}}\\[\tunedvs]
&\qquad\qquad\col{13${}_N$} \quad 0 = \CC{B_1\dots B_N}{A}{(\mu\nu|} \tilted{M^{A|\gamma)}} \\[\tunedvs]
&\qquad\qquad\col{14${}_N$} \quad 0 = \Coeff{AB_1\dots B_{N-1}} \left( \tilted{M^{B[\mu|} M^{A|\nu]}{}_{:B}} + \left(\mathrm{deg} P - 1\right) \init{p{}^{\rho[\mu|}}\straight{F^{A}{}_\rho{}^{|\nu]}} \right) \\[\tunedvs]
&\qquad\qquad\col{15${}_N$} \quad 0 = \CC{B_1\dots \widehat{B_J}\dots B_{N+1}}{B_J}{\mu\nu} - \CC{\vphantom{\widehat{B_J}}B_1\dots B_{N}}{B_{N+1}}{\mu\nu} \qquad \text{for} \quad J = 1 \dots N+1  \\[\tunedvs]
&\qquad\qquad\col{16${}_N$} \quad 0 = N\cdot(N+1) \left(\mathrm{deg} P-1\right) \Coeff{AB_1\dots B_N} \init{p^{\rho(\mu|}} \straight{F^A{}_\rho{}^{|\nu)}} + N \CC{B_1\dots B_N}{A}{(\mu|} \tilted{M^{A|\nu)}} + 2N \CC{B_1\dots B_N}{A}{\alpha(\mu|} \tilted{M^{A|\nu)}{}_{,\alpha}} + (N-2) \CC{B_1\dots B_{N-1}}{B_N}{\mu\nu} \\[\tunedvs]
&\qquad\qquad\col{17${}_N$} \quad 0 = (N+2)\cdot (N+1) \left(\mathrm{deg} P-1\right) \Coeff{AB_1\dots B_{N+1}} \left(\init{p^{\mu\nu}}\straight{E^A{}_\nu} - \init{p^{\mu\nu}{}_{,\gamma}} \straight{F^A{}_\nu{}^\gamma}\right) - (N+1)^2 \Coeff{A(B_1\dots B_N} \tilted{M^{A\mu}{}_{:B_{N+1})}} - (N+1) \CC{B_1\dots B_{N+1}}{A}{} \tilted{M^{A\mu}} \\[0pt]
&\qquad\qquad\qquad\qquad\qquad  - (N+1) \CC{B_1\dots B_{N+1}}{A}{\alpha} \tilted{M^{A\mu}{}_{,\alpha}} - (N+1) \CC{B_1\dots B_{N+1}}{A}{\alpha\beta} \tilted{M^{A\mu}{}_{,\alpha\beta}} - \sum_{K=1}^{N+1} \CC{B_1\dots \widehat{B_K} \dots B_{N+1}}{B_K}{\mu} + 2 \left(\partial_\gamma \CC{B_1\dots B_N}{B_{N+1}}{\gamma\mu}\right) \\
&\qquad\qquad\col{18${}_N$} \quad 0 = \CC{A}{B}{\mu_1\dots \mu_N} - \sum_{K=0}^{\textrm{\tiny max}} (-1)^{K+N} \binom{K+N}{K} \left(\partial^K_{\alpha_1\dots \alpha_K} \CC{B}{A}{\alpha_1\dots \alpha_K \mu_1\dots \mu_N}\right) \\
&\qquad\qquad\col{19${}_N$} \quad 0 = \sum_{K=0}^{\textrm{\tiny max}} \binom{K+N}{K} \CC{B}{A}{\alpha_1\dots \alpha_K(\mu_1\dots \mu_{N}|} \tilted{M^{A|\mu_{N+1})}{}_{,\alpha_1\dots\alpha_K}} + \sum_{K=0}^{\textrm{\tiny max}} (-1)^{K+N} \binom{K+N+1}{N+1} \left( \partial^K_{\alpha_1\dots\alpha_K} \CC{}{B}{\alpha_1\dots\alpha_K\mu_1\dots \mu_{N+1}} \right) \\
&\qquad\qquad\col{20${}_{N\textrm{\tiny even}}$} \quad 0 = \sum_{K=N}^{\textrm{\tiny max}} \sum_{J=N+1}^{K+1} (-1)^J \binom{K}{J-1} \binom{J}{N} \partial^{J-N}_{\alpha_1\dots\alpha_{J-N}} \left( \CC{}{A}{\beta_J\dots\beta_K(\alpha_1\dots\alpha_{J-N}\mu_1\dots\mu_{N-1}|} \tilted{M^{A|\mu_N)}{}_{,\beta_J\dots\beta_K}} \right) \\[\tunedvs]
&\qquad\qquad \col{21${}_{N\textrm{\tiny odd}}$} \quad 0 = 2 \sum_{K=N-1}^{\textrm{\tiny max}} \binom{K}{N-1} \CC{}{A}{\beta_N\dots\beta_K(\mu_1\dots\mu_{N-1}|} \tilted{M^{A|\mu_N)}{}_{,\beta_N\dots\beta_K}} \\[0pt]
&\qquad\qquad\qquad\qquad\qquad - \sum_{K=N}^{\textrm{\tiny max}} \sum_{J=N+1}^{K+1} (-1)^J \binom{K}{J-1} \binom{J}{N} \partial^{J-N}_{\alpha_1\dots\alpha_{J-N}} \left( \CC{}{A}{\beta_J\dots\beta_K(\alpha_1\dots\alpha_{J-N}\mu_1\dots\mu_{N-1}|} \tilted{M^{A|\mu_N)}{}_{,\beta_J\dots\beta_K}} \right)
\end{align*}
\end{small}
\end{table}
\end{turnpage}
\end{document}